\documentclass[prb,showpacs,preprintnumbers,amsmath,amssymb,twocolumn,superscriptaddress]{revtex4}

\usepackage{graphicx}
\usepackage{dcolumn}
\usepackage{bm}
\usepackage{color}

\begin{document}

\title{
Equilibrium Surface Current and Role of U(1) Symmetry:\\
sum rule and surface perturbations
}

\author{Yasuhiro Tada}

\affiliation{Institute for Solid State Physics, The University of Tokyo,
Kashiwa 277-8581, Japan}

\newcommand{\vecc}[1]{\mbox{\boldmath $#1$}}

\begin{abstract}
We discuss 
effects of surface perturbations 
on equilibrium surface currents which contribute
to orbital magnetization and orbital angular momentum 
in systems without time reversal symmetry.
We show that,
in a U(1) particle number conserving system, 
disorder and other perturbations
at a surface
do not affect the equilibrium surface current and corresponding
orbital magnetization due to a sum rule
which is analogous to Luttinger's theorem. 
On the other hand, 
for a superfluid,
the sum rule is no longer applicable and hence
the surface mass current and corresponding orbital angular
momentum
can depend on details of a surface.
\end{abstract}

\pacs{Valid PACS appear here}

\maketitle

\section{introduction}
\label{sec:intro}
The orbital magnetization (OM) and orbital angular momentum (OAM) 
are one of the
most fundamental physical quantities in condensed matter physics,
which arise in systems without time reversal symmetry
due to equilibrium circulating currents.
In a continuum (non-lattice) 
system which does not break translational symmetry 
in bulk regions, 
an equilibrium circulating current flows only near its boundary.
While in a lattice system, a circulating current is also 
possible around each atom in addition to a surface current.
In spite of their physical importance, however,
OM/OAM and associated equilibrium 
surface currents are not well understood.
Indeed, the OM arising from surface currents 
are usually not taken into account but only locally circulating currents
around atoms are included, when one calculates total magnetization of a system
in the density functional theory~\cite{book:Martin2004}.
This would be partly because it is difficult to appropriately treat
surface currents theoretically and also effects of surface currents on
the total magnetization are expected to be small in conventional
ferromagnets such as Fe and Ni.

However,
contributions of surface currents to OM/OAM would become 
important
in several interesting systems
such as topological insulators/superconductors without
time reversal symmetry and fractional quantum Hall systems
where there are chiral edge modes
\cite{pap:Hasan2010,pap:Qi2011,book:Wen2004}.
In the last decade,
many theoretical studies for OM have been proposed 
for band insulators and metals
including topological systems, 
and they give a beautiful formula for calculating OM.
\cite{pap:Gat2003a,pap:Gat2003b,
pap:Xiao2005,pap:Thonhauser2005,pap:Ceresoli2006,pap:Shi2007,
pap:Resta2010,pap:Thonhauser2011,pap:Xiao2010,
pap:Matsumoto2011,pap:Chen2011,pap:Chen2012}.
This formula involves only the Bloch wavefunctions which are bulk
properties of a system, and hence, 
the OM and associated surface currents can be regarded as bulk quantities.
These results are in good agreement with our general expectations that
magnetism is a bulk property which is independent of surface
details in real materials.
Conceptually, they can be thought as
a nice realization
of the bulk-surface correspondence in a general sense that 
bulk properties are determined by surface physics and vice versa.
At the same time,
however, one may naively expect that surface currents could be
affected by perturbations near surfaces
which should exist in real materials,
such as surface disorder,
deformation of Wannier functions, local inversion symmetry breaking, 
weak screening of the Coulomb interaction, and so on.

A partial answer to this fundamental question can be obtained 
by following the derivations of the formula for OM.
The formula has been derived in three different ways;
(i) direct calculations of circulating currents for trivial band insulators
in the presence of boundaries~\cite{pap:Thonhauser2005,pap:Ceresoli2006}, 
(ii) semi-classical wavepacket approximations
~\cite{pap:Xiao2005,pap:Xiao2010,pap:Matsumoto2011},
and (iii) taking derivative of free energy with respect to
magnetic fields under the periodic boundary condition
~\cite{pap:Shi2007,pap:Chen2011}.
Following the derivation (i), one could find that the surface
currents are independent of details of the boundaries when
the ground state is a simple band insulator. 
Based on this, it was argued that
effects of surface perturbations on OM in such simple
insulators are irrelevant~\cite{pap:Chen2012}.
Although the discussion presented in the derivation (i) cannot be
applied to other systems,
the same formula was obtained by the derivations (ii) and (iii).
The surface currents are found to be independent of gradient of
surface potentials
within the semi-classical wavepacket approximations in the derivation (ii). 
In this approximation,
however, the length scale of surface potentials should be much longer than
the wavepacket size.
In the derivation (iii), although 
the OM is given as a bulk quantity by its definition 
for systems with periodic boundary conditions,
connections to a surface current which exists in a realistic 
finite size system are unclear~\cite{pap:Shi2007,pap:Chen2011}.
Therefore, in spite of the surprisingly convincing agreements between the three
derivations,
it is still not clear why the OM and corresponding surface currents
are given as bulk quantities which are independent of surface details.

In contrast to non-superconducting systems discussed above,
surface perturbations do become relevant for
OAM in chiral superfluids which break time reversal symmetry.
The OAM in chiral superfluids has been a long-standing issue
and attracting much interest since the discovery of $^3$He A-phase
~\cite{pap:Leggett1975,book:Vollhart1990,
book:Volovik2003,book:Leggett2006,pap:Ishikawa1980,pap:Mermin1980,
pap:Kita1998,pap:Goryo1998,
pap:Furusaki2001,pap:Stone2004,pap:Mizushima2008,
pap:Sauls2011,pap:Bradlyn2012,pap:Shitade2014,pap:Hoyos2014,
pap:Tsutsumi2014,pap:Volovik2015,pap:Tada2015,pap:Huang2014}.
Effects of surface roughness have been studied 
in the context of $^3$He A-phase and Sr$_2$RuO$_4$
~\cite{pap:Maeno2003,pap:Maeno2012},  
and it was theoretically argued that
the surface mass currents and resulting OAM in chiral superfluids 
depend on surface roughness in weak coupling regions
\cite{pap:Sauls2011,pap:Ashby2009,pap:Nagato2011}.
In case of domain walls between chiral BCS superfluids with
opposite chiralities, the boundary currents strongly depend on
difference in U(1) phases of order parameters
\cite{pap:Tsutsumi2014,pap:Volovik2015}.
Besides, when the surface is sharp,
net surface currents vanish for higher order pairing 
states such as $d+id, f+if$-wave BCS superfluids due to hidden depairing
effects which exist even for clean surfaces~\cite{pap:Tada2015,pap:Huang2014}. 
These results suggest that, in contrast to U(1) symmetric
systems, surface currents and OAM in superfluids
with broken time reversal symmetry
depend on surface details and are not bulk quantities.
However, 
physical origins of the reduction of the surface currents 
and difference between
the superfluids and U(1) symmetric systems
have not been well understood.
Besides, 
while surface perturbations are known to be
relevant in the weak coupling BCS states where 
Cooper pairs are extended in space, 
they have not been
discussed so far for the strong coupling BEC states
where Cooper pairs are strongly bounded~\cite{pap:Read2000}. 
For such tightly bounded pairing states, one may naively
expect that the surface currents are robust against surface perturbations.

In this paper, 
in order to clarify the different behaviors of the surface currents
and corresponding OM/OAM in systems with or without U(1) symmetry,
we discuss effects of surface perturbations from a general point of view.
Based on a sum rule argument which is analogous to Luttinger's theorem 
~\cite{pap:Luttinger1960,pap:LuttingerWard1960,pap:Oshikawa1997,
pap:Oshikawa2000,
pap:Dzyaloshinskii2003}
and numerical calculations,
we show that surface currents are robust against
surface perturbations in general U(1) symmetric systems.
On the other hand, 
for superfluids without time reversal symmetry,
the sum rule is no longer applicable and
surface currents can depend on surface perturbations such as 
surface roughness. 
Especially, it is shown that
surface currents are suppressed in chiral superfluids on a lattice 
even for strong coupling BEC states.

This paper organizes as follows.
In Sec.~\ref{sec:U(1)}, we discuss surface currents in
U(1) symmetric systems.
We show a sum rule for the surface current in general systems
and confirm it
by numerical calculations for a concrete model.
The OM formula is revisited based on the sum rule argument.
Then, surface currents in U(1) broken systems are examined in
Sec~\ref{sec:nonU(1)}.
Finally, we summarize this paper in Sec.~\ref{sec:summary}

\section{system with U(1) symmetry}
\label{sec:U(1)}
\subsection{sum rule argument}
\label{sec:sum_rule}
\begin{figure}
\begin{center}
\includegraphics[width=0.6\hsize,height=0.4\hsize]{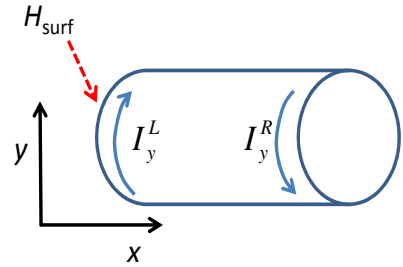}
\caption{
Schematic picture of the
two-dimensional cylinder system.
}
\label{fig:cylinder}
\end{center}
\end{figure}

In this section, we discuss surface currents in 
U(1) symmetric systems.
For simplicity, we consider 2-dimensional lattice models
whose size is $N_x\times N_y$
with the open boundary condition for $x$-direction and
the periodic boundary condition for $y$-direction as shown
in Fig.~\ref{fig:cylinder}.
Our argument holds also for continuum models 
as discussed in Appendix A.
We assume that there is no time reversal symmetry
and there can exist equilibrium surface currents for the models
considered.
Our Hamiltonian reads in general
\begin{align}
H&=\sum c^{\dagger}_{xl}(k_y)K_{xl,x'l'}(k_y)
c_{x'l'}(k_y)+H_{\rm int}+H_{\rm surf}
\label{eq:H}
\end{align}
where $c_{xl}(k_y)$ is an annihilation operator of electrons at 
position $x$, with wavenumber along $y$-direction $k_y$ and
other quantum number $l$ such as orbitals and spins.
The matrix $\hat{K}$ describes one-particle Hamiltonian.
We assume that the interaction $H_{\rm int}$ is simply
short-range
density-density interactions or on-site interactions so that it commutes with
the total particle density operator, $[n_i,H_{\rm int}]=0$.
$H_{\rm surf}$ is the surface perturbation term which is
finite only near the surface and zero otherwise,
and is also assumed to satisfy $[n_i,H_{\rm sruf}]=0$.
Therefore, the U(1) current operator is simply given by
the kinetic term only through the continuity equation in the present study.

In this geometry, the equilibrium surface current at the left (right)
surface is given by
\begin{align}
I_y^{L(R)}=\frac{1}{N_y}\sum_{xx'\in S_L(S_R)}v_{y;xl,x'l'}(k_y)
\langle c^{\dagger}_{xl}(k_y)
c_{x'l'}(k_y)\rangle,
\end{align}
in unit of the electron charge $e$.
$S_{L(R)}$ is a region only near the left (right)
surface whose width is
much smaller than the system width $N_x$.
Since effects of a surface should not propagate deep into bulk regions,
we can define such regions in general systems~\cite{com:j}.
$\hat{v}_y$ is a velocity matrix given by $\hat{v}_y(k_y)
=\partial \hat{K}(k_y)/\partial k_y$.
In the absence of $H_{\rm surf}$
it is obvious that $I_y^{\rm tot}\equiv I_y^L+I_y^R=0$ 
 under a natural assumption
that the Hamiltonian~\eqref{eq:H} has inversion symmetry
in the $x$-direction, $x\leftrightarrow -x$.
Since current density which contributes to the surface current
is localized only around the surface and it vanishes in bulk regions out of
$S_{L,R}$~\cite{com:j},
the total surface current $I_y^{\rm tot}$ is written as
\begin{align}
I_y^{\rm tot}&=\frac{1}{N_y}\sum_{xx'\in \mbox{all sites}}v_{y;xl,x'l'}(k_y)
\langle c^{\dagger}_{xl}(k_y)
c_{x'l'}(k_y)\rangle\nonumber \\
&=\frac{1}{N_y}\sum_{k_y}{\rm tr}\bigl[\hat{v}_y(k_y)\hat{G}(k_y,i\omega)
e^{i\omega 0^+}\bigr].
\label{eq:Itot}
\end{align}
Here, $\hat{G}(k)=\hat{G}(k_y,i\omega)
=-\langle\langle c(k_y)c^{\dagger}(k_y)\rangle\rangle
(i\omega)$
is the matrix Matsubara Green's function, and
trace describes summation over all the indices, $x,l$ and $\omega$.
We note that, although off diagonal elements become finite
if the system breaks some symmetries together with time reversal
symmetry such as spin rotation symmetry in
a ferromagnetic state, 
our discussion holds in a parallel way by introducing order parameters
into $\hat{G}$ and modifying the definition of $\hat{v}_y$
appropriately.
For example, if the translational symmetry along the $y$-direction
is preserved, only possible off diagonal elements in the Green's function
are of the form
$\langle c_{x'l'}(k_y)c^{\dagger}_{xl}(k_y)\rangle$ which
have already been included in the above expression.
If the translational symmetry along the $y$-direction is broken,
although
off diagonal elements with different momenta $k_y$ and
$k_y+Q_y$ with an ordering vector $Q_y$ become finite,
Eq. \eqref{eq:Itot} is still valid after
modifying $\hat{v}_y(k_y)$ to $\tilde{v}_y(k_y)
={\rm diag}(\hat{v}_y(k_y),\hat{v}_y(k_y+Q_y))$ where $k_y$ belongs to
the reduced Brillouin zone.

Now we turn on $H_{\rm surf}$ which is
finite only near the left surface and zero in all the other
regions including the right surface.
Our fundamental assumption is that $I_y^R$ is unchanged
by the left surface perturbation $H_{\rm surf}$ as long as
the width of the cylinder is large enough,
which means that if $I_y^{\rm tot}$ changes, it is totally 
attributed to change in $I_y^L$.
We also assume that $H_{\rm surf}$ is translationally
symmetric along the surface. 
In case of surface disorder, $H_{\rm surf}$ is 
still translationally symmetric on average, 
for which the disorder-averaged Green's 
function is diagonal with respect to $k_y$.
Then, all the effects of such perturbations in one-particle quantities
can be incorporated into the (averaged)
selfenergy matrix $\hat{\Sigma}(k)$.
Effects of $H_{\rm int}$ are also described by the selfenergy, and
if the system is in broken symmetry states, 
we include the corresponding order parameter matrix in $\hat{\Sigma}(k)$.
The Green's function $\hat{G}(k)$ is now written as a matrix inverse of 
$[i\omega-\hat{K}(k_y)-\hat{\Sigma}(k)]$.
Therefore,
\begin{align}
I_y^{\rm tot}&=\frac{1}{N_y}\sum_{k_y}{\rm tr}\Bigl[
\frac{\partial \hat{K}(k_y)}{\partial k_y}
\hat{G}(k)\Bigr]\notag\\
&=-\frac{1}{N_y}\sum_{k_y}{\rm tr}\Bigl[
\frac{\partial \hat{G}^{-1}}{\partial k_y}
\hat{G}\Bigr]
-\frac{1}{N_y}\sum_{k_y}{\rm tr}\Bigl[
\frac{\partial \hat{\Sigma}}{\partial k_y}
\hat{G}\Bigr]\notag\\
&=-\frac{1}{2\pi}{\rm tr}\log \Bigl[\frac{
\hat{G}(k_y=\pi,i\omega)}{\hat{G}(k_y=-\pi,i\omega)}\Bigr]
-\frac{1}{N_y}\sum_{k_y}{\rm tr}\Bigl[
\frac{\partial \hat{\Sigma}}{\partial k_y}
\hat{G}\Bigr].
\label{eq:Itot2}
\end{align}
It is clear that the first term vanishes.
(If the translational symmetry is broken, $k_y=\pm \pi$
should be replaced by appropriate boundary $k$-vectors of
the reduced Brillouin zone.)
Second term also vanishes because of the Luttinger-Ward identity
\cite{pap:Luttinger1960,pap:LuttingerWard1960,pap:Dzyaloshinskii2003,
com:LW},
\begin{align}
\frac{1}{N_y}\sum_{k_y}{\rm tr}\Bigl[
\frac{\partial \hat{\Sigma}(k)}{\partial k_y}
\hat{G}(k)\Bigr]=0.
\label{eq:LW}
\end{align}
Although
the Luttinger-Ward identity may be violated in Mott insulators
~\cite{pap:Rosch2007,pap:Dave2013},
it holds in general gapless/gapped Fermi liquids where
there is no zero (pole) in the Green's function
(selfenergy), and also in weakly disordered systems~\cite{com:LW}.
As long as the Luttinger-Ward identity is satisfied,
$I_y^{\rm tot}=I_y^L+I_y^R=0$ even in the presence
of the perturbations around the left surface,
which means that $I_y^L$ is unchanged although the current 
density at each site could be affected by $H_{\rm surf}$.

There is a nice analogy between 
the conservation of surface currents in the presence of surface 
perturbations and the well known conservation law,
the Luttinger's theorem~\cite{pap:Luttinger1960,pap:LuttingerWard1960,
pap:Oshikawa1997,pap:Oshikawa2000,pap:Dzyaloshinskii2003}. 
The Luttinger's theorem claims that, though
shape of a Fermi surface can be changed by interactions,
its total volume is unchanged.
Here, although a real space profile of U(1) surface current density $j_i$
can be modified by the surface perturbations,
its sum around the surface $I=\sum_{i\in {\rm surf}}j_i$ is unchanged.
In this view point, 
the conservation of the U(1) surface current 
in the presence of surface 
perturbations is understood as a result of non-trivial cancellations
of changes in $j_i$ at each site.
This sum rule argues that surface currents and corresponding OM are
bulk quantities which are independent of surface details
as suggested in the modern theories on 
OM~\cite{pap:Xiao2005,pap:Thonhauser2005,pap:Ceresoli2006,
pap:Shi2007,
pap:Resta2010,pap:Thonhauser2011,pap:Xiao2010,
pap:Matsumoto2011,pap:Chen2011,pap:Chen2012}, 
and this supports existence of the bulk-surface correspondence 
for these quantities.
We note that,
the sum rule is useful not only for developments in understanding 
fundamental aspects of surface currents, but also for practical calculations of
them.
Although realistic surface potentials would be complicated in general,
one can use a particular surface potential which is suitable for computing 
them,
such as a hard wall potential and a sufficiently smooth potential.
The sum rule guarantees that the calculation results of the surface current
and OM are independent of the potential used.
Indeed, the known formula for OM can be reproduced by calculating surface
current contributions
and also contributions from locally circulating currents
under a sufficiently smooth surface potential,
as will be discussed in Sec.~\ref{sec:OMformula}.

Up to now, we have not considered external 
or spontaneously generated magnetic fields.
In the presence of an applied magnetic field parallel to the $z$-direction,
the hopping integrals acquire phase factors corresponding to the flux
configuration by the Peierls substitution,
and the Brillouin zone is reduced to a magnetic Brillouin zone.
Even in this case, by appropriately modifying the definition of $\hat{v}(k_y)$,
we can still derive a sum rule for a surface current in the same way.
The surface current under a magnetic field for a time reversal symmetric system
is related to the 
Landau diamagnetism.
It has been known that 
``bulk approaches" and ``surface approaches" are equivalent
for the Landau diamagnetism;
it can be calculated either by
derivative of free energy with respect to the magnetic field under periodic
boundary conditions~\cite{pap:Landau1930,pap:Peierls1933,pap:Fukuyama1971}
or by computing a surface current in the presence of boundaries
~\cite{pap:Kubo1964,pap:Ohtaka1973,pap:Ishikawa1999}, 
and these two approaches consistently give the same results.
Experimentally, the skipping orbits which would be responsible for
the surface current
have been observed in 
surface impedance measurements for many metals
~\cite{pap:Nee1968,book:Abrikosov}.
Our sum rule argument would 
give a new understanding on
the known equivalence between the two theoretical approaches.

Finally,
it is noted that,
the sum rule can hold 
since the U(1) charges of the electrons are the well defined
unique value $e$ in the systems with U(1) symmetry.
If U(1) symmetry is absent, however, 
particle sectors with charge $+e$ and hole sectors with $-e$
are mixed and the surface currents will not generally be conserved,
as will be discussed in Sec.~\ref{sec:nonU(1)}.

\subsection{Bloch-Bohm's theorem}
\label{sec:BlochBohm}
In this section, we discuss relations of our sum rule 
to the Bloch-Bohm's theorem.
The theorem states that net current should vanish
in the ground states~\cite{pap:Bohm1949,pap:Vignale1995,pap:Ohashi1996,
pap:Kusama1999}.
Here, we reexamine this theorem and point out that
(i) it does not hold when surface currents are concerned and (ii)
it needs some modifications when spontaneous symmetry breaking
is involved even for currents running in bulk regions.

We consider a general Hamiltonian as in Eq.~\eqref{eq:H} without
symmetry breaking fields in a finite size
cylinder $L\times L$ where open (periodic) boundary condition is imposed for
$x(y)$-direction, and denote the ground state wavefunction as
$|0_{L}\rangle$.
Since the system size is finite, there is no spontaneous symmetry breaking
and the ground state wavefunction
preserves the symmetries 
of the Hamiltonian including the time reversal symmetry
if it is contained in $H$.
When the Hamiltonian has time reversal symmetry, 
it is trivial that the total current $I^{\rm tot}_y$
vanishes because it is odd under the time reversal symmetry.
In the following, we mainly discuss $H$ with time reversal symmetry
and $H$ without time reversal symmetry will be touched on briefly.
For time reversal symmetric $H$, 
we introduce external fields $\lambda H_{\rm ex}
=\lambda\sum \Delta_{il,jl'}c^{\dagger}_{il}c_{jl'}$ which break
time reversal symmetry by assuming that the state
$|0_L\rangle$ of $H$ has corresponding instability, 
where $\lambda$ is a dimensionless constant
which should be taken as $\lambda \rightarrow0$ in the end.
Although we focus on U(1) symmetric external fields in this section,
effects of U(1) symmetry breaking fields will be discussed in 
Sec.~\ref{sec:nonU(1)}.

Following \textcite{pap:Bohm1949},
we consider a variational state $|\theta_{L,\lambda}\rangle
=U_{\theta}|0_{L,\lambda}\rangle$ where $|0_{L,\lambda}\rangle$
is the ground state of $H_{\lambda}=H+\lambda H_{\rm ex}$.
The unitary operator is defined as $U_{\theta}=\exp[i\theta
\sum_{j}yn_{j}]$~\cite{pap:LSM1961,pap:Oshikawa1997,pap:Oshikawa2000}, 
and $\theta=2\pi n/L, (n=0,\pm1,\cdots)$ is
required so that $U_{\theta}$ is a well-defined operator.
This requirement can easily be understood from the first quantization
form of $U_{\theta}$; the twist operator for the wavefunction
$\Psi(\vecc{r}_1,\cdots,\vecc{r}_N)=\Psi(\vecc{r}_1+L,\cdots,\vecc{r}_N+L)$ 
is given by
$U_{\theta}=\exp[i\theta\sum_j y_j]$. In order for
$\Psi'(\vecc{r}_1,\cdots,\vecc{r}_N)=(U_{\theta}\Psi)
(\vecc{r}_1,\cdots,\vecc{r}_N)$  to satisfy the periodic boundary condition,
$\theta=2\pi n/L$ is needed.
If $\theta$ were not an integer multiple of $2\pi/L$,
the wavefunction $(U_{\theta}\Psi)$ is no longer an element of the
domain of the Hamiltonian.
This is essentially comes from the well-known fact that 
position operators cannot be well-defined on a torus, although it is sometimes
missed even by experts~\cite{pap:Carruthers1968}.
Similarly in the second quantization formalism, 
$c_{jl}'=U^{\dagger}_{\theta}c_jU_{\theta}
=e^{-i\theta j}c_{jl}$ obeys the periodic boundary condition
when $\theta=2\pi n/L$ is satisfied.
Although this fundamental point has been missing in most of 
the previous studies concerning Bloch-Bohm's theorem,
this is important especially for discussing possible net surface currents.

We then evaluate energy difference between $|\theta_{L,\lambda}\rangle$
and $|0_{L,\lambda}\rangle$,
\begin{align}
\delta E_{L,\lambda}
&=\langle \theta_{L,\lambda}|H_{\lambda}|\theta_{L,\lambda}\rangle 
-\langle 0_{L,\lambda}|H_{\lambda}|0_{L,\lambda}\rangle \notag\\
&=\sum (\cos[\theta(y_i-y_j)]-1)\langle 0_{L,\lambda}|
c_{il}^{\dagger}K_{il,jl'}c_{jl'}|0_{L,\lambda}\rangle\notag\\
&\quad +\lambda\sum (\cos[\theta(y_i-y_j)]-1)
\langle 0_{L,\lambda}|
c_{il}^{\dagger}\Delta_{il,jl'}c_{jl'}|0_{L,\lambda}\rangle\notag\\
&\quad +\sum i\sin[\theta(y_i-y_j)]\langle 0_{L,\lambda}|
c_{il}^{\dagger}K_{il,jl'}c_{jl'}|0_{L,\lambda}\rangle\notag\\
&\quad +\lambda\sum i\sin[\theta(y_i-y_j)]
\langle 0_{L,\lambda}|
c_{il}^{\dagger}\Delta_{il,jl'}c_{jl'}|0_{L,\lambda}\rangle.
\end{align}
It is well known that
the Lieb-Schultz-Mattis twist operator $U_{\theta}$
is not helpful for higher dimensions 
other than one-dimension, in order to construct 
a variational state with low energy excitations. 
Indeed,
since $\cos[\theta(y_i-y_j)]-1\sim O(1/L^2)$ in the present model with
a finite hopping range $l_t$ which is much shorter than $L$,
the $\cos[\theta(y_i-y_j)]$-terms are obviously $O(1)$ for large $L$ limit.
In general $d$-dimensional system of an isotropic size $L$ for all directions,
these terms are $O(L^{d-2})$.
Therefore, 
the $\sin[\theta(y_i-y_j)]$-terms could become dominant for large $L$,
if at least one of them is of order $L$ or larger, namely
$\sum i\sin[\theta(y_i-y_j)]\langle c^{\dagger}_iKc_j\rangle\sim O(L)$
or $\sum i\sin[\theta(y_i-y_j)]\langle c^{\dagger}_i
\Delta c_j\rangle\sim O(L)$.
By expanding $\sin[\cdots]$, we see that one of the $\sin[\cdots]$-terms
is simply the total current running in the whole system, 
$\theta\langle 0_{L,\lambda}
|\sum_i j_{yi}|0_{L,\lambda}\rangle$ in the leading order of $\theta\ll1$,
where $j_{yi}$ is current density.
The original Bloch-Bohm's theorem states that,
since we can choose either $\theta=2\pi/L$ or $\theta=-2\pi/L$,
in order for $\delta E_{L,\lambda}$ to be non-negative,
$\langle 0_{L,\lambda}
|\sum_i j_{yi}|0_{L,\lambda}\rangle$ must be zero in the thermodynamic limit
$L\rightarrow \infty$.
However,
the Bloch-Bohm's argument is based on an implicit assumption that 
the expectation value of the total current is 
$\langle \sum_ij_{yi}\rangle \sim O(L^2)$, and also 
the other sin-term is negligible, $\sum i(y_i-y_j)\langle c^{\dagger}_i
\Delta c_j\rangle\sim o(L)$.
At the same tiem,
we also should pay attention to the $\cos[\cdots]$-terms which are 
$O(1)$.
In the case of surface currents without any current density in the bulk, 
contributions only come from the 
surface regions $S_{L,R}\sim O(1)$, and therefore 
$\theta\langle \sum_ij_{yi}\rangle \sim (1/L)\times L=O(1)$ and 
it becomes the same order as the $\cos$-terms.
In $d$-dimensional systems, 
$\theta\langle \sum_ij_{yi}\rangle \sim (1/L)\times L^{d-1}=O(L^{d-2})$
and it competes with the $\cos$-terms as well,
for which we cannot immediately conclude that finite $\langle 0_{L,\lambda}
|\sum_i j_{yi}|0_{L,\lambda}\rangle\sim O(L^{d-1})$ 
contradicts with $\delta E_{L,\lambda}$.
Hence, {\it
the Bloch-Bohm's argument is not applicable when surface currents
are concerned,}
although it might give an upper bound for the net surface currents.
On the other hand, the sum rule argument in the previous section
claims that surface currents should be exactly canceled between opposite
surfaces in a cylinder.

The above observation is applicable either with or without
the symmetry breaking fields.
In the following, we discuss some subtleties related to the external
fields and the thermodynamic limit, which have also been overlooked
in the previous studies. 
Here, we focus on possibilities of total currents 
of order $L^2$
($L^d$ in $d$-dimensions), which are bulk currents but not surface currents.
Firstly, we consider a case where $H$ itself does not
have time-reversal symmetry without symmetry breaking fields,
although such a Hamiltonian would be artificial.
In this case, we could keep $L$ to be some large but finite values,
so that the total current term becomes dominant in $\delta E_{L,\lambda=0}$.
Then, we can safely apply the original Bloch-Bohm's argument to conclude that
total current of order $L^2$ must vanish for sufficiently large $L$.

However, if we include the symmetry breaking fields for 
time reversal symmetric $H$,
we have extra terms in $\delta E_{L,\lambda}$ arising from 
$\lambda H_{\rm ex}$ and the resulting $O(L)$ term in $\delta E_{L,\lambda}$
is not determined by the total current only,
$\delta E_{L,\lambda}\sim (\pm2\pi/L) \langle \sum [j_{yi}+
\lambda(\cdots)]\rangle$ where $\lambda(\cdots)$ represents
$\lambda i(y_i-y_j)c^{\dagger}_i\Delta c_{j}$.
At this point, what is forbidden by the Bloch-Blohm's argument is that
the above expectation value is some finite value of order $L^2$ for large $L$.
We do not know, however, which contribution in the bracket
becomes dominant for given $L,\lambda$,
although the second term may be smaller for sufficiently small $\lambda$.
Furthermore, since we are interested in the 
spontaneously symmetry broken states, we should take the limit of
vanishing external fields $\lambda\rightarrow0$. 
When the system is defined on
$x,y=-L/2+1,\cdots,L/2$ with an even $L$, we consider
\begin{align}
\delta \varepsilon_n\equiv\lim_{\lambda \downarrow0}\lim_{L\uparrow \infty}
\frac{\delta E_{L,\lambda}}{L^n}.
\end{align}
The contribution arising from $\lambda H_{\rm ex}$ will vanish as
$\lambda\rightarrow 0$ due to the prefactor $\lambda$ in front of
$H_{\rm ex}$,
as long as the expectation value $\lim_{L\uparrow\infty}
\sum(2\pi/L^{n+1})
\langle0_{L,\lambda}|c^{\dagger}\Delta c|0_{L,\lambda}\rangle$ is not
singular at $\lambda=0$.
Because we have assumed that time reversal symmetry is broken and 
the corresponding order parameter is finite in the thermodynamic limit,
this term should be some constant which is independent of $\lambda$
in the limit $\lambda\rightarrow0$, and therefore we can safely take
the limit.
It is noted that difference in  the thermodynamic energy density 
$\delta \varepsilon_2$ vanishes in this limit, while 
$\delta \varepsilon_1$  can be negative if
the total surface current per volume is $O(1)$ for finite $L$
and is non-zero
after taking the limit.
However, we should be careful about meaning of 
the possible non-zero $\delta\varepsilon_1$.
In the thermodynamic limit, quantum states for fixed $\lambda$
may be defined as 
\begin{align}
\omega_0(\cdot)&=\lim_{L\uparrow \infty}
\langle 0_{L,\lambda}|\cdot|0_{L,\lambda}\rangle,\\
\omega_{\theta}(\cdot)&=\lim_{L\uparrow \infty}
\langle \theta_{L,\lambda}|\cdot|\theta_{L,\lambda}\rangle
\end{align}
for
local operators~\cite{book:Emch1972}.
There are some subtleties in the thermodynamic limit, where there is no
local operator describing the total current density for the whole
system corresponding to
$\delta \varepsilon_1\sim \langle \sum j_{yi}\rangle/L^2$.
In this case,
the two thermodynamic states $\omega_0,\omega_{\theta}$ become 
identical; 
\begin{align}
\omega_0(A)=
\omega_{\theta}(A)
\end{align}
for
any local operator $A$.
For example, $\omega_{\theta}(c^{\dagger}_{il}c_{jl'})
=\lim_{L\uparrow\infty} 
e^{i\theta(y_i-y_j)}\langle 0_{L,\lambda}|c^{\dagger}_{il}c_{jl'}|
0_{L,\lambda}\rangle=
\omega_{0}(c^{\dagger}_{il}c_{jl'})$
because $|y_i-y_j|<l_t$ is finite and $\lim_{L\uparrow\infty}
e^{i\theta(y_i-y_j)}=1$.
(Note that expectation values of U(1) breaking local operators 
vanish trivially for both states.)
This is true even when the two states are orthogonal for finite $L$,
and the two orthogonal states can 
converge to a single state in the thermodynamic limit.

Such a behavior arises from global nature of the variational state
$|\theta_{L,\lambda}\rangle$.
The two states $|0_{L,\lambda}\rangle,|\theta_{L,\lambda}\rangle$,
are almost identical locally and their difference appears as a sum of
these tiny local differences.
In the present system, 
to obtain different states in the thermodynamic limit,
we need to restrict the twist only for a finite support such as $D=\{(x,y)|
1\leq x-\bar{x},y-\bar{y}\leq L'\}$ where $(\bar{x},\bar{y})$ is
an arbitrary site.
We then define a new variational state $|\theta'_{L,\lambda}\rangle=U_{\theta}'
|0_{L,\lambda}\rangle$, $U_{\theta}'=\exp[i\theta'\sum'y_jn_j]$ where
$\theta'=2\pi/L'$ and the summation is taken only 
for the above finite domain $D$~\cite{pap:Oshikawa1997}.
It is noted that, in order for $U_{\theta}'$ to be well-defined,
$L'$ should be $L/L'=$integer.
When we take the thermodynamic limit, we keep $L'$ constant but
increase $L$ only.
Nevertheless, we can take $L'$ to be much larger than the finite hopping range
of the model $l_t\ll L'$.
Then, the energy difference for $\theta'=\pm 2\pi/L'$ is 
$\delta E_{L,\lambda}=\pm
(2\pi/L')\sum'\langle 0_{L,\lambda}|j_{yi}+\lambda(\cdots)|0_{L,\lambda}\rangle
+O(l_t/L')$.
By taking the limit, the energy difference for $\theta=\pm 2\pi/L'$ becomes
\begin{align}
\lim_{\lambda \downarrow0}\lim_{L\uparrow \infty}
\delta E_{L,\lambda}=\pm\frac{2\pi}{L'} 
\lim_{\lambda \downarrow0}\lim_{L\uparrow \infty}
\langle 0_{L,\lambda}|\sum_i' j_{yi}|0_{L,\lambda}\rangle+ O(L'^0),
\label{eq:dE'}
\end{align}
where the $O(L'^0)=O(1)$ term come from the cos-term.
Note that the first term does not describe total current 
of the system, but it corresponds to current running within the finite region 
$D$.
Here, 
we recall that the ground state of $H_{\lambda}$
in the thermodynamic limit is defined so that
$\omega_0(A^{\dagger}[H_{\lambda},A])\geq 0$ 
is satisfied for any local operator $A$.
If we take $A=U_{\theta}'$, this means 
\begin{align}
\lim_{\lambda \downarrow0}
\omega_0(U_{\theta}'^{\dagger}H_{\lambda}U_{\theta}'
-H_{\lambda})=\lim_{\lambda \downarrow0}\lim_{L\uparrow \infty}
\delta E_{L,\lambda}\geq 0.
\label{eq:dE''}
\end{align}
By comparing Eqs.~\eqref{eq:dE'} and ~\eqref{eq:dE''}, 
we conclude that 
the current running in the domain $D$ cannot be of order $O(L'^2)$
for sufficiently large $L'\gg l_t$.
This statement is a bit stronger than the original Bloch-Bohm's argument,
since the position of $D$ characterized by the site $(\bar{x},\bar{y})$
can be arbitrary in the infinite system.
Hence, we conclude that {\it there is no macroscopic flow anywhere in a
thermodynamic system.}
We will see that this statement holds for U(1) symmetry broken systems
as well  in Sec.~\ref{sec:nonU(1)}.
The only remaining possibilities would be
locally circulating currents in an atomic scale
and surface currents which are of smaller order in $L'$ as discussed before.

\subsection{numerical simulation}
\label{sec:Hah}
In order to confirm the sum rule argument,
we perform numerical calculations of surface currents
in a simple toy model defined on a square lattice cylinder. 
We impose the open (periodic) boundary condition 
for the $x(y)$-direction. 
As typical examples of surface perturbations,
we investigate effects of surface roughness and surface potentials.
We consider the following
Hamiltonian
\begin{align}
H_{\rm AH}&=H_{\rm AH0}+H_{\rm surf},
\label{eq:Hah}\\
H_{\rm AH0}&=\sum_{\rm NN}[-tc^{\dagger}_{i1}c_{j1}
+tc^{\dagger}_{i2}c_{j2}+t'e^{i\theta_{ij}}c^{\dagger}_{i1}c_{j2}
+({\rm h.c.})]\notag \\
&\quad +\sum M[n_{i1}-n_{i2}],\notag\\
H_{\rm surf}&=\sum_{} V_{il}n_{il},\notag
\end{align}
where $e^{i\theta_{i,i\pm\hat{x}}}=\mp i,
e^{i\theta_{i,i\pm\hat{y}}}=\pm 1$, and $V_{il}$ is the surface perturbation.
This model is a spinless version of the 
Bernevig-Hughes-Zhang model, and does not 
have time reversal symmetry~\cite{pap:Hasan2010,pap:Qi2011}.
In the present calculations, $V_{il}$ is
finite only at the left surface sites $i=(x=1,y)$.
We study two cases: (i)
$V_{i1,2}$ take real random values in $[-V_0,V_0]$ in case of surface 
roughness,
and (ii) $V_{i1,2}$ are constant, $(V_{i1},V_{i2})=(V_0,0)$,
as a particular realization of surface potentials.
Since the $t'$-term is simply a part of kinetic term which arises from
spin-orbit interaction
in the original Bernevig-Hughes-Zhang model, the current density operator for 
the $\mu=x,y$-direction in the present model is given by
\begin{align}
j_{\mu i}&=-it[c^{\dagger}_{i1}c_{i+\hat{\mu}1}-c^{\dagger}_{i+\hat{\mu}1}
c_{i1}]
+it[c^{\dagger}_{i2}c_{i+\hat{\mu}2}-c^{\dagger}_{i+\hat{\mu}2}
c_{i2}]
\notag\\
&\quad+it'e^{i\theta_{ii+\hat{\mu}}}
[c^{\dagger}_{i1}c_{i+\hat{\mu}2}+c^{\dagger}_{i+\hat{\mu}1}
c_{i2}]+({\rm h.c.}).
\label{eq:jah}
\end{align}
For simplicity, we fix $(t',M)=(0.2t,-3t)$ as an example,
and filling is $n=n_1+n_2=0.7$ for which the system is in a
metallic anomalous Hall
state.
Temperature is fixed at $T=0$.
System size is $N_x\times N_y=80\times20$ for the case (i),
and numerical results are checked for other systems sizes.
For the case (ii), with use of Fourier transformation for the
$y$-direction, larger system sizes are examined.

\begin{figure}
\begin{center}
\includegraphics[width=0.8\hsize,height=0.5\hsize]{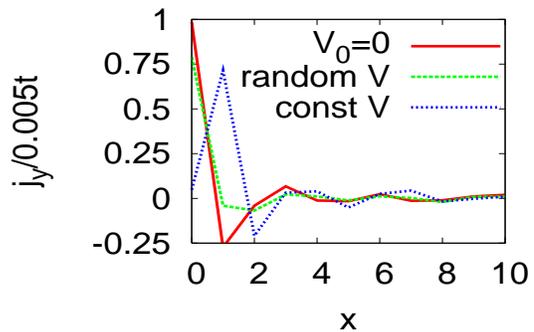}
\caption{
Current density $j_y$ near the left surface for $V_0=0$ (red),
random potential with $V_0=2t$ (green), and constant potential with $V_0=2t$.
}
\label{fig:jah}
\end{center}
\end{figure}

\begin{figure}
\begin{tabular}{lr}
\hspace{-0.5cm}
\begin{minipage}{0.6\hsize}
\begin{center}
\includegraphics[width=0.8\hsize,height=0.6\hsize]{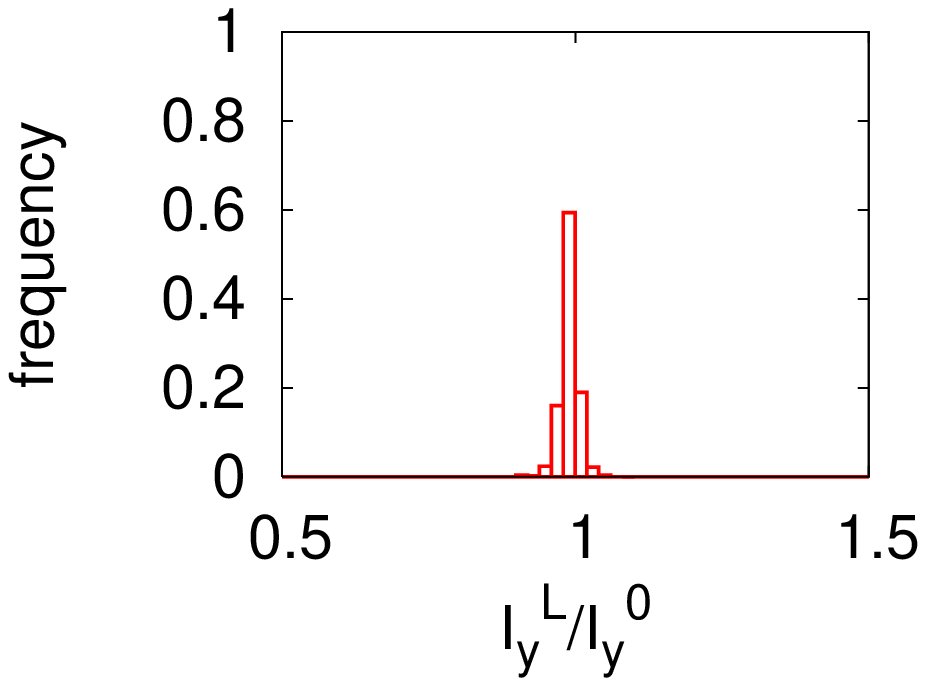}
\end{center}
\end{minipage}
\hspace{-1.0cm}
\vspace{0.5cm}
\begin{minipage}{0.6\hsize}
\begin{center}
\includegraphics[width=0.8\hsize,height=0.6\hsize]{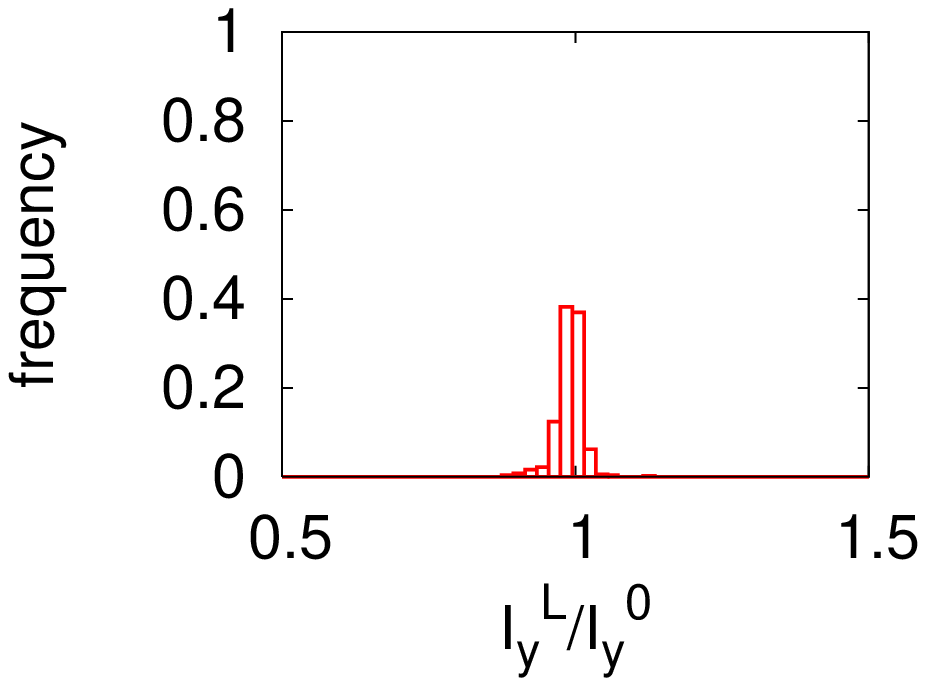}
\end{center}
\end{minipage}
\end{tabular}
\caption{
Normalized histogram of $I_{y}^L$ for
disorder potentials $V_0=0.5t$ (left panel) and $V_0=2t$ (right panel).
}
\label{fig:hist}
\end{figure}

In Fig.~\ref{fig:jah},
we show the current densities $j_{yi}$ for the random potential and 
constant potential at a large
$V_0=2t\gg t'=0.2t$ together with $j_{yi}$ for $V_0=0$
(which we denote $j_{yi}^0$ hereafter).
We see that
$j_{xi}$ vanishes everywhere in the system and $I^R_y$ is unchanged
by the left surface potentials.
For the random potential, we take a disorder average $\langle j_{yi}
\rangle_{\rm av}$ and then
take an average of them along the $y$-direction,
$(1/N_y)\sum_{y=1}^{N_y}\langle j_{yi}
\rangle_{\rm av}$.
It is seen that, although there are some small oscillations in the bulk region
due to metallicity in the present model, $j_{yi}$ is localized around the
surfaces.
The current densities are modified from $j_{yi}^0$ around the left surface
$x=0$.
However, we find that, both for the random potential and constant potential, 
the left surface current $I_y^L=(1/N_y)\sum_{i\in{\rm surf}}
j_{yi}$ is unchanged from $I_y^0=(1/N_y)\sum_{i\in{\rm surf}}
j_{yi}^0$, which is confirmed for several different system sizes.
Conservation of the surface currents were also seen in the previous study
for a constant surface potential in an insulating state~\cite{pap:Chen2012}.
In the case of disorder potential, 
distribution of $I_y^L$ for
different configurations of
the potential is well localized around its mean value $I_y^0$
as shown in Fig.~\ref{fig:hist}.
It is noted that the distribution of $I_y^L$
gets broader if we introduce inter-orbital random potentials 
$\sum [V_{i12}c_{i1}^{\dagger}c_{i2}+({\rm h.c.})]$ in addition to
the intra-orbital potential (not shown).
Although 
finite size effects become rather large in this case, 
the average surface current is still unchanged by the surface disorder.
We have performed similar calculations for other realizations of 
surface potentials, and confirmed that the surface current is
unchanged by them.
These numerical calculations indeed support the sum rule argument
in Sec.~\ref{sec:sum_rule}.

\subsection{Revisit of Orbital Magnetization Formula}
\label{sec:OMformula}
\begin{figure}
\begin{center}
\includegraphics[width=0.5\hsize,height=0.4\hsize]{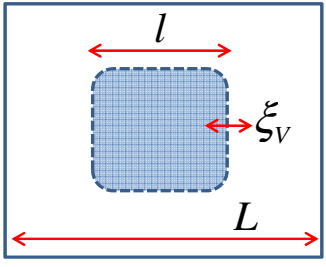}
\caption{
Schematic picture of the system with the confinement potential $V$
which is zero in the shaded region and infinite in the white region.
$V$ changes in a length scale $\xi_V$ which is much longer than
the lattice constant $a$.
These length scales satisfy
$a\ll \xi_V\ll l \ll L$.
}
\label{fig:appB}
\end{center}
\end{figure}

As mentioned in Sec.~\ref{sec:sum_rule},
the sum rule is helpful not only for basic understanding 
of surface currents but also
for practical calculations of them.
Here, based on an observation of the sum rule,
we rederive the formula of orbital magnetization $\vecc{M}$ 
at $T=0$~\cite{pap:Xiao2005,pap:Thonhauser2005,pap:Ceresoli2006,pap:Shi2007,
pap:Resta2010,pap:Thonhauser2011,pap:Xiao2010,
pap:Matsumoto2011,pap:Chen2011,pap:Chen2012}
for
a thermodynamically large but finite size system.
We consider a non-interacting model of size $L\times L$
with the periodic boundary condition, whose Hamiltonian
is expressed in a Wannier function basis as 
$H_0=\int_{L^2}d^2x \psi^{\dagger}{\mathcal K}\psi
=\sum_{ijll'}c^{\dagger}_{il}\hat{H}_0c_{jl'}$.
The chemical potential has been included in $H_0$.
Then, we introduce a confinement potential, $H_{\rm surf}=
\sum_{i}V_{i}n_i$, which smoothly varies in space and is zero 
inside a region $l\times l$ while infinitely large outside.
The length scales can be taken as 
\begin{align}
a\ll \xi_V\ll l \ll L 
\end{align}
where
$a$ is the lattice constant and $\xi_V$ characterizes the spatial
variation length of $V$.
The system configuration is schematically shown in Fig.~\ref{fig:appB}.
Although such a confinement potential would not be a
realistic surface potential, the sum rule
guarantees that the surface current is equivalent to that 
with a realistic confinement potential.
It is noted that, as long as the length scale of the confinement
potential is much shorter than the system size $\xi_V\ll l$,
the locally circulating current around each atom in the bulk is not affected by
the potential.

Generally,
the paramagnetic current density in unit of $e$ and orbital magnetization 
in a finite system with boundaries are given by
~\cite{pap:Eschrig1985,pap:Vignale1987,pap:Higuchi1997},
\begin{align}
\vecc{j}(\vecc{r})&=\frac{-i}{2m}\bigl[\psi^{\dagger}\nabla\psi
-\nabla\psi^{\dagger}\psi\bigr],\label{eq:jp}\\
M_{z}\times {\rm vol}
&=\frac{1}{2}\int_{\rm finite} d^2x[\vecc{r}\times\vecc{j}(\vecc{r})]_z
\notag\\
&=\frac{1}{2}\sum_{i}\int_{v_i}d^2x (\vecc{r}-\vecc{R}_i)\times\vecc{j}
(\vecc{r})|_z\notag\\
&\qquad\qquad +\frac{1}{2}\sum_{i}\vecc{R}_i\times \int_{v_i}d^2x\vecc{j}
(\vecc{r})|_z,
\label{eq:M}
\end{align}
where
$v_i$ is a unit cell with its center position $\vecc{R}_i$.
The first term in $M_{z}$ arises from locally circulating current,
while the second term is due to the surface current.

We first consider the surface contribution.
In the presence of the smooth confinement potential,
derivative expansion in the Wigner representation would be
legitimate~\cite{book:Volovik2003,pap:Kubo1964},
where the site index $i$ can be considered as a continuum variable
in the length scale $\xi_V\gg a$.
The Green's function is approximated up to the lowest order 
with respect to derivative of the confinement potential by,
\begin{align}
\hat{G}(\vecc{X},k)&=\hat{G}_0\Bigl[
1+\frac{i}{2}\frac{\partial \hat{G}_0^{-1}}{\partial X_{\mu}}
\frac{\partial \hat{G}_0}{\partial k_{\mu}}-
\frac{i}{2}\frac{\partial \hat{G}_0^{-1}}{\partial k_{\mu}}
\frac{\partial \hat{G}_0}{\partial X_{\mu}}
\Bigr]\notag\\
&=\hat{G}_0\Bigl[
1+\frac{i}{2}\frac{\partial \hat{G}_0^{-1}}{\partial X_{\mu}}
\frac{\partial \hat{G}_0}{\partial k_{\mu}}
-\frac{i}{2}\hat{G}_0^{-1}\frac{\partial \hat{G}_0}{\partial k_{\mu}}
\frac{\partial \hat{G}_0^{-1}}{\partial X_{\mu}}\hat{G}_0
\Bigr],
\end{align}
where $\hat{G}_0(\vecc{X},k)$ is the matrix inverse of 
$[i\omega-\hat{H}_0(\vecc{k})-V(\vecc{X})]=\hat{G}_0^{-1}$ with respect to
the indices $l,l'$.
$\vecc{X}=(\vecc{x}_1+\vecc{x}_2)/2$ is the center of mass coordinate
and $\vecc{k}$ is a wavevector corresponding to the relative
coordinate $\vecc{x}_1-\vecc{x}_2$.
In the above, we have used 
$0=\partial (\hat{G}_0^{-1}\hat{G}_0)/\partial k_{\mu}=
(\partial \hat{G}_0^{-1}/\partial k_{\mu})\hat{G}_0+\hat{G}_0^{-1}
(\partial \hat{G}_0/\partial k_{\mu})$.
The surface current along the $x$-direction is given by
\begin{align}
I_x&=\int_{X_y^1}^{X_y^2} dX_y {\rm tr}\frac{\partial \hat{H}_0}{\partial k_x}
\hat{G}\notag\\
&=\int dX_y \frac{-i}{2}\frac{\partial V}{\partial X_y}
{\rm tr}\hat{G}_0
\Bigl[
\frac{\partial \hat{G}_0^{-1}}{\partial k_x}\frac{\partial \hat{G}_0}{\partial k_y}-
\frac{\partial \hat{G}_0^{-1}}{\partial k_y}\frac{\partial \hat{G}_0}{\partial k_x}
\Bigr],
\end{align}
where $\partial V/\partial X_x=0$ near $[010]$ surface 
and $\int dX_y$ is restricted around the surface where
$V(X_y^1)=0$ and $V(X_y^2)=\infty$.
tr represents summation over all the indices other than $\vecc{X}$.
By using $G_0(\vecc{X},k)=\sum_n|u_{kn}\rangle\langle u_{kn}|/(i\omega-
\varepsilon_{kn}-V(\vecc{X}))$, we can perform integral over $\omega$
and simplify the expression as was done in ~\textcite{pap:Chen2011}
to obtain
\begin{align}
I_x&=-\int dX_y \frac{d V}{d X_y}\frac{1}{L^2}\sum_{kn}
f(\varepsilon_{kn}+V)\Omega_{kn}^{z},\notag\\
&=-\frac{1}{L^2}\sum_{kn} \int^{\infty}_{\varepsilon_{kn}} 
d\varepsilon f(\varepsilon)\Omega_{kn}^{z}\notag\\
&=\frac{1}{L^2}\sum_{\varepsilon_{kn}<0}
\varepsilon_{kn}\Omega_{kn}^{z},\\
\Omega_{kn}^{z}&=i\epsilon_{\mu\nu}\int_{L^2}d^2x
\frac{\partial u_{kn}^{\ast}}{\partial k_{\mu}}
\frac{\partial u_{kn}}{\partial k_{\nu}},
\end{align}
where $f$ is the Fermi distribution function at $T=0$.
This is nothing but the surface current evaluated within
the quasi-classical wavepacket theory in the previous studies
~\cite{pap:Xiao2005,pap:Xiao2010,pap:Matsumoto2011}.
The orbital magnetization 
arising from the surface current is,
\begin{align}
M_z^{\rm surf}&=\frac{1}{L^2}
\sum_{\varepsilon_{kn}<0} \varepsilon_{kn}\Omega_{kn}^{z}.
\end{align}

Next, we consider the bulk contribution,
$M_{z}^{\rm bulk}=\sum_i(1/2l^2)
\int_{v_i} d^2x[(\vecc{r}_q-\vecc{R}_i)\times\vecc{j}]_z
\simeq N_{l}(1/2l^2)
\int_{v_0} d^2x[\vecc{r}_q\times\vecc{j}]_z$
where $v_0$ is the unit cell with its center $\vecc{R}_0=0$
and $N_{l}$ is the number of unit cells inside 
the confinement potential.
$l^2=v_0N_l$ is satisfied.
Here, the position $\vecc{r}$ has been replaced by
$r_q^{\mu}=\sin (qr^{\mu})/q$ where $q=2\pi/L$
which is consistent with the periodic boundary
condition. 
However,
in order to calculate this contribution,
we can safely approximate $\vecc{r}_q$ as $\vecc{r}_q\simeq \vecc{r}
+O(v_0^{1/2}/L)$ in the integrand.
Besides, we can simply neglect effects of the confinement potential and
approximate the Green's function as
$G(\vecc{x}_1,\vecc{x}_2)\simeq G_0(\vecc{x}_1,\vecc{x}_2)$
for $\vecc{x}_1,\vecc{x}_2\in v_0$, 
Then,
\begin{align}
M_{z}^{\rm bulk}=\frac{-i}{4mv_0}
\int_{v_0} d^2x[\vecc{r}\times(\nabla_1-\nabla_2)G_0(\vecc{x}_1,\vecc{x}_2)
|_{x_1=x_2}]_z
\end{align}
can now be directly calculated, e.g. by the Green's function 
method in the first principles calculations~\cite{book:Martin2004}.
Alternatively,
we can also use $G_0(\vecc{x}_1,\vecc{x}_2)=
\sum_{kn}\phi_{kn}(\vecc{x}_1)\phi^{\ast}_{kn}(\vecc{x}_2)
/(i\omega-\varepsilon_{kn})$ for $\vecc{x}_1,\vecc{x}_2\in v_0$ and
express this contribution in terms of Bloch functions
by a formal calculation.
By denoting $p_{\nu}=-i\partial_{\nu}/m$ and $H_k=e^{-ikr}
{\mathcal K}e^{ikr}$, 
we have
\begin{align}
J&\equiv\epsilon_{\mu\nu}\int_{v_0} d^2x\phi_{kn}^{\ast}x_{\mu}p_{\nu}
\phi_{kn}\notag\\
&=\epsilon_{\mu\nu}
\sum_{k'n'}\int_{v_0} d^2x \phi_{kn}^{\ast}
\left(e^{ikr}i\partial_{k_\mu}e^{-ikr}
\right)\phi_{k'n'}\notag\\
&\qquad \times \int_{L^2} d^2x\phi_{k'n'}^{\ast}p_{\nu}\phi_{kn}\notag\\
&=\epsilon_{\mu\nu}
\sum_{k'n'}-\delta_{kk'}\delta_{nn'}i\partial_{k_{\mu}'}
\int_{v_0}d^2x u_{k'n'}^{\ast}e^{-ik'r}p_{\nu}e^{ikr}u_{kn}\notag\\
&\quad -\epsilon_{\mu\nu}
\int_{v_0}d^2x e^{-ikr}i\partial_{k_{\mu}}u_{kn}^{\ast}(\vecc{x})\notag\\
&\qquad \times\int_{L^2}d^2x' \sum_{k'n'}
\phi_{k'n'}(\vecc{x})\phi_{k'n'}^{\ast}(\vecc{x}')
p_{\nu}\phi_{kn}(\vecc{x}')\notag\\
&=-J-2i\epsilon_{\mu\nu}
\int_{v_0}d^2x i\partial_{k_{\mu}}u_{kn}^{\ast}
\frac{\partial H_k}{\partial k_{\nu}}u_{kn}.
\end{align}
This leads to
\begin{align}
J&=-i\epsilon_{\mu\nu}
\int_{v_0}d^2x \frac{\partial u_{kn}^{\ast}}{\partial k_{\mu}}
\frac{\partial H_k}{\partial k_{\nu}}u_{kn}\notag\\
&=i\epsilon_{\mu\nu}
\int_{v_0}d^2x \frac{\partial u_{kn}^{\ast}}{\partial k_{\mu}}
(H_k-\varepsilon_{kn})\frac{\partial u_{kn}}{\partial k_{\nu}}\notag\\
&\quad -i\epsilon_{\mu\nu}
\int_{v_0}d^2x \frac{\partial u_{kn}^{\ast}}{\partial k_{\mu}}
\frac{\partial \varepsilon_{kn}}{\partial k_{\nu}}u_{kn}.
\end{align}
The second term in the above expression vanishes after taking
$\sum_{kn}$ as
\begin{align}
J'&\equiv i\sum_{\varepsilon_{kn}<0}\epsilon_{\mu\nu}
\int_{v_0}d^2x \frac{\partial u_{kn}^{\ast}}{\partial k_{\mu}}
\frac{\partial \varepsilon_{kn}}{\partial k_{\nu}}u_{kn}\notag\\
&=i\sum_{\varepsilon_{kn}<0}\sum_{k'}\epsilon_{\mu\nu}\delta_{kk'}
\int_{v_0}d^2x \frac{\partial u_{k'n}^{\ast}}{\partial k_{\mu}'}u_{kn}
\frac{\partial \varepsilon_{kn}}{\partial k_{\nu}}\notag\\
&=-i\sum_{\varepsilon_{kn}<0}\epsilon_{\mu\nu}
\frac{\partial^2 \varepsilon_{kn}}{\partial k_{\mu}\partial k_{\nu}}
-J'.
\end{align}
Therefore, we obtain
\begin{align}
M_z^{\rm bulk}
&=\frac{i\epsilon_{\mu\nu}}{2v_0}\sum_{\varepsilon_{kn}<0}
\int_{v_0}d^2x \frac{\partial u_{kn}^{\ast}}{\partial k_{\mu}}
(H_k-\varepsilon_{kn})\frac{\partial u_{kn}}{\partial k_{\nu}},
\end{align}
which agrees with the previous works~\cite{pap:Xiao2005,pap:Xiao2010,
pap:Matsumoto2011}.
By collecting the two contributions, $M_{z}^{\rm surf}$ and $M_{z}^{\rm bulk}$,
we finally end up with the OM formula
\begin{align}
M_z&=M_{z}^{\rm surf}+M_{z}^{\rm bulk}\notag\\
&=\frac{i\epsilon_{\mu\nu}}{2}
\sum_{n}\int_{\varepsilon_{kn}<0} \frac{d^2k}{(2\pi)^2}
\int_{L^2}d^2x \frac{\partial u_{kn}^{\ast}}{\partial k_{\mu}}
(H_k+\varepsilon_{kn})\frac{\partial u_{kn}}{\partial k_{\nu}}.
\end{align}
We note that, since $M_z^{\rm bulk}$ is independent of the chemical potential
in insulators,
we can reproduce the Streda formula
in terms of
the surface current only~\cite{pap:Streda1982,book:MacDonald1995},
\begin{align}
\frac{\partial M_z}{\partial \mu}
&=\frac{\partial M_z^{\rm surf}}{\partial \mu}
=-\frac{1}{2\pi}\nu,\\
\nu&=
\sum_{n:{\rm occ}}\frac{1}{2\pi}\int_{\rm BZ} d^2k
\Omega_{kn}^z,
\end{align}
where $\mu$ is in the gap.

Finally, let us briefly discuss relations of the present results in
bounded systems to the previous calculations in periodic systems
~\cite{pap:Xiao2005,pap:Shi2007,pap:Xiao2010,pap:Chen2011}.
In the following, for simplicity, 
we consider electromagnetic coupling up to the first order in 
$\vecc{B}=(0,0,B)$
at a fixed gauge, in order to discuss OM at zero field.
For a uniform magnetic field $\vecc{A}=(1/2)\vecc{B}\times \vecc{r}$
in a bounded finite size system,
\begin{align}
H_{\rm EM}=-
\int_{\rm finite}d^2x\vecc{j}\cdot \vecc{A}
=-\Bigl(\frac{1}{2}\int_{\rm finite}d^2x \vecc{r}\times \vecc{j}
\Bigr) \cdot \vecc{B}
\label{eq:EM}
\end{align}
holds as an operator identity, where $\vecc{j}$ is the
paramagnetic current Eq.~\eqref{eq:jp}
~\cite{pap:Higuchi1997}.
Expectation value of the above integrand 
is not uniform in space for both expressions, and
the surface current contribution is localized around the surface.
Nevertheless, the integrated energy
$\langle H_{\rm EM}\rangle$ is independent of surface conditions,
since the OM is a bulk quantity as implied by the sum rule argument.
Then, it is quite natural that the total energy 
including $\langle H_{\rm EM}\rangle$ of the bounded system
is equivalent to that in a periodic system of the same volume
with the uniform magnetic field $\vecc{B}$.
This should be true from a macroscopic point of view that
total energy of a system which is an extensive quantity 
does not depend on boundary conditions in the leading order of
the system size.
Therefore, OM calculated by derivative of the free energy with respect
to $\vecc{B}$ in the periodic system is
equivalent to that in the bounded system computed either from 
Eq.~\eqref{eq:M} or from derivative of Eq.~\eqref{eq:EM}.
However, from a microscopic point of view,
the coincidence of these two quantities
is not a priori guaranteed,
because OM in the periodic system
is defined only by the derivative of the free energy
but is not given by an expectation value of 
an OM operator such as Eq.~\eqref{eq:M} since it
is ill-defined under periodic boundary conditions.
In this sense, 
the sum rule of surface current density gives a microscopic
basis for the macroscopic equivalence of total energies 
under different boundary conditions.

\section{system without U(1) symmetry}
\label{sec:nonU(1)}
\subsection{general argument}
In this section, we consider superfluids without time reversal symmetry.
As noted briefly in the previous section,
in case of systems without U(1) symmetry,
we cannot apply sum rule arguments on robustness of surface
currents against surface perturbations.
The main difficulty arises from the modification of 
the velocity matrix in the Green's function formalism,
\begin{align}
\hat{v}_y(k_y)\rightarrow
\tilde{v}_y(k_y)&=\frac{1}{2}
\left[
\begin{array}{cc}
\hat{v}_y(k_y) & 0\\
0 & -\hat{v}_y^T(-k_y)
\end{array}
\right]\notag\\
&=\frac{1}{2}Q\frac{\partial \tilde{K}}{\partial k_y},
\end{align}
where $Q={\rm diag}(1,-1)$ and
$\tilde{K}(k_y)={\rm diag}(\hat{K}(k_y),-\hat{K}^T(-k_y))$
in the Nambu space.
The charge matrix $Q$ is not a unit matrix,
since the charge carried by the particles is assigned to be $+1$ while 
it is $-1$ for the holes.
Correspondingly,
$I_y^{\rm tot}$ is not simply given by a simple form
as in Eq.~\eqref{eq:Itot2}.
For this case, we cannot simply perform the summation over $k_y$
in $I_y^{\rm tot}$, and cannot obtain a sum rule for the surface current.
Instead, we can consider a related quantity
$\tilde{I}_y^{\rm tot}$
which is given by
\begin{align}
\tilde{I}_y^{\rm tot}
&\equiv \frac{1}{N_y}\sum_{k_y}{\rm tr}\Bigl[
Q^{-1}\tilde{v}(k_y)\tilde{G}(k)\Bigr]\notag\\
&=-\frac{1}{N_y}\sum_{k_y}{\rm tr}
\Bigl[\frac{\partial \tilde{G}^{-1}}{\partial k_y}
\tilde{G}\Bigr]
-\frac{1}{N_y}\sum_{k_y}{\rm tr}
\Bigl[\frac{\partial \tilde{\Sigma}}{\partial k_y}
\tilde{G}\Bigr].
\end{align}
Here, $\tilde{G}$ and $\tilde{\Sigma}$ are respectively the Green's function
and the selfenergy in the Nambu representation, and trace includes
summation over the Nambu space.
While $I^{\rm tot}$ describes difference between 
particle-like contribution and hole-like contribution,
$\tilde{I}^{\rm tot}$ is related to an equal-weighted sum of these
contributions and is conserved in the presence of surface perturbations
due to the sum rule.
This is analogous to the problem of Fermi surface volumes
where each of them and hence difference among them
are not conserved in general, 
while their total sum is unchanged by interactions
as stated in Luttinger's theorem.
We note that,
for general SU($N$) currents such as spin currents and orbital currents, 
corresponding charge matrices are not the unit matrix in 
the spin/orbital space and there would be 
off diagonal matrix elements 
in the velocity matrix such as spin-orbit coupling
and inter-orbital hybridization.
Similarly to the U(1) broken case,
we will have the same problem and cannot apply sum rule
arguments to those cases.

Similar difficulty arises in the Bloch-Bohm's argument
for superfluidity.
Although it is not helpful for surface currents as was discussed in
Sec.~\ref{sec:BlochBohm}, we briefly discuss it in superfluids
for a comparison with U(1) symmetric systems, focusing on
macroscopic bulk currents.
The largest difference between U(1) symmetric systems and
U(1) broken systems comes from the external field $\lambda H_{\rm ex}$
under twist by $U_{\theta}=\exp[i\theta\sum y_jn_j]$.
When $\lambda H_{\rm ex}=\lambda \sum \Delta_{il,jl'}c^{\dagger}_{il}
c_{jl'}^{\dagger}+({\rm h.c.})$,
it is transformed as
\begin{align}
U^{\dagger}_{\theta}\lambda H_{\rm ex}U_{\theta}
=\lambda \sum e^{i\theta (y_i+y_j)}\Delta_{il,jl'}c^{\dagger}_{il}
c_{jl'}^{\dagger}+({\rm h.c.}).
\end{align}
If we evaluate energy difference between the ground state 
$|0_{L,\lambda}\rangle$ and a variational state
$|\theta_{L,\lambda}\rangle=U_{\theta}|0_{L,\lambda}\rangle$,
we obtain, in the leading order of $\theta=2\pi n/L$,
\begin{align}
\delta E_{L,\lambda}&=\langle \theta_{L,\lambda}|H_{\lambda}|\theta_{L,\lambda}\rangle 
-\langle 0_{L,\lambda}|H_{\lambda}|0_{L,\lambda}\rangle\notag\\
&=\lambda\sum (\cos[\theta(y_i+y_j)]-1)\langle 0_{L,\lambda}|
c_{il}^{\dagger}\Delta_{il,jl'}c_{jl'}^{\dagger}|0_{L,\lambda}\rangle\notag\\
&\quad +\lambda\sum i\sin[\theta(y_i+y_j)]\langle 0_{L,\lambda}|
c_{il}^{\dagger}\Delta_{il,jl'}c_{jl'}^{\dagger}|0_{L,\lambda}\rangle\notag\\
&\quad +({\rm h.c.}).
\end{align}
Since $y_i+y_j$ can be of order $L$, we cannot Tayler expand 
$\cos[\cdots]/\sin[\cdots]$ and neglect higher order terms
in $(y_i+y_j)/L$.
On the contrary, above two terms will be of order $L^2$, and because
$(\cos[\theta(y_i+y_j)]-1)=-2\sin^2[\theta(y_i+y_j)/2]<0$
while $\sin[\theta(y_i+y_j)]$ is oscillating in sign,
the first term would become dominant.
Indeed, the latter term will vanish if $|0_{L,\lambda}\rangle$ is
translationlly invariant in a long distance scale $\sim 1/\theta$.
Sign of the first term can be evaluated, once
we simply assume $\langle 0_{L,\lambda}|\lambda H_{\rm ex}|
0_{L,\lambda}\rangle<0$, which is reasonable for spontaneous symmetry
breaking.
This assumption 
is a variant of the statement that external magnetic fields
parallel to the magnetic moment lower the total energy in conventional 
ferromagnets.
Indeed, $\langle 0_{L,\lambda}|\lambda H_{\rm ex}|
0_{L,\lambda}\rangle$ is a part of condensation energy of superfluidity,
and therefore should be negative when the finite size system has 
instability towards the corresponding superfluidity.
If the above assumption really holds, we see that 
$\delta E_{L,\lambda}\sim \lambda\times o(L^2)>0$
in the leading order of $L$ 
by noting that
$-2\sin^2[\theta(y_i+y_j)]<0$ can be approximated by 
a negative constant of order unity.

However, similarly to U(1) symmetric systems, the variational state 
$\omega_{\theta}$ becomes identical to $\omega_0$ 
in the limit $\lambda\rightarrow0$,
and $\lim_{\lambda\downarrow0}
\lim_{L\uparrow\infty}\delta E_{L,\lambda}/L^2$ vanishes because of the
prefactor $\lambda$ in front of $H_{\rm ex}$.
Therefore, we introduce the other variational state $|\theta'_{L,\lambda}
\rangle$ which is twisted only in a finite domain $D$.
By repeating the same calculation, we find that the leading contribution in
$\delta E_{L,\lambda}$ comes from $\lambda H_{\rm ex}$ which is $O(L'^2)$,
and next leading $O(L')$-contribution is the total current term 
if $\langle\sum_{i\in D} j_{iy}\rangle\sim O(L'^2)$. 
For the twist only in $D$, however, the former $O(L'^2)$ term will vanish
in the limit $\lim_{\lambda\downarrow0}\lim_{L\uparrow\infty}
\delta E_{L,\lambda}$, and
we obtain Eq.~\eqref{eq:dE'} as in U(1) symmetric systems.
(As mentioned for U(1) symmetric systems, 
$\lim_{L\uparrow\infty}
\lambda\langle \sum' e^{i\theta(y_i+y_j)}c^{\dagger}_i\Delta 
c^{\dagger}_j\rangle$
would not be singular at $\lambda=0$ and will vanish for $\lambda\rightarrow0$.)Therefore, we arrive at the same statement as for U(1) symmetric systems that
macroscopic currents are not allowed anywhere in the ground states of 
superfluids in the
thermodynamic limit.
This statement also holds in the presence of static magnetic
fields in superconductors and macroscopic
supercurrents flowing in the bulk are not allowed at equilibrium.
Therefore, if a Fulde-Ferrell state or a helical state with non-zero
center of mass momenta of Cooper pairs
is realized, there should be some counter-propagating currents which
compensate the macroscopic supercurrents.
For example, it was pointed out that in the helical states 
in noncentrosymmetric superconductors under magnetic fields,
supercurrents are canceled by magnetization currents 
and there are no currents in the thermodynamic
limit~\cite{pap:Kaur2005,pap:Yip2005}.

The Bloch-Bohm's argument can predict
vanishing macroscopic currents which is proportional to domain volumes, 
while the Green's function approach is less helpful for U(1) broken systems.
However, neither of them can exclude possible net currents
due to incomplete
cancellations of surface currents.

\subsection{numerical simulation}
Although the sum rule discussions cannot be applied to superfluids,
it is still possible that the surface mass current is robust
against surface perturbations by some other reasons.
In order to investigate this,
we examine numerically surface currents in two simple 
models, a non-chiral $p$-wave superfluid based on the model~\eqref{eq:Hah}
and a chiral $p$-wave superfluid.
We focus on neutral fermions and do not consider Meissner effects
in the present study.
Temperature is fixed at $T=0$.

\subsubsection{non-chiral $p$-wave superfluid}
Firstly, we consider a non-chiral $p$-wave superfluid
based on the model~\eqref{eq:Hah} in order to examine
how U(1) symmetry breaking modifies the previous results
in Sec. \ref{sec:Hah}.
The Hamiltonian is 
\begin{align}
H_{{\rm AHSF}}&=H_{{\rm AH0}}+H_{\rm surf}
-g\sum_{}n_{il}n_{i+\hat{y}l},\\
H_{\rm surf}&=\sum V_{ill'}c^{\dagger}_{il}c_{il'},\notag
\label{eq:Hahs}
\end{align}
where $g$ is an attractive interaction for $p_y$-wave superfluidity.
As in Sec.~\ref{sec:Hah}, we again consider two particular 
examples of surface perturbations, a random potential and
a constant potential.
In case of disorder potential,
inter-orbital surface potentials $V_{i12}=V_{i21}^{\ast}$
are introduced in addition to the intra-orbital potentials $V_{i11,22}
=V_{i1,2}$.
This Hamiltonian is one of the simplest models to discuss
intra-orbital superfluidity in the model~\eqref{eq:Hah},
and this superfluidity itself does not break time-reversal symmetry.
We perform mean field calculations of the superfluidity.
The mean field calculations are performed
for each disorder configuration in the case of disorder potential, 
which is repeated
until averaged physical values become converged.
It is noted that,
in the cylinder geometry where the periodic boundary condition
is imposed for the $y$-direction,
there is no zero-energy Andreev bound state at the surfaces, 
which makes numerical calculations rather stable.
The system size mostly used for the random potential 
is $N_x\times N_y=40\times20$
and results are qualitatively unchanged for other sizes 
up to $N_x\times N_y=60\times20$.
For the constant potential, similarly to the previous section,
we can perform Fourier transformation for the $y$-direction
and study larger sizes.

\begin{figure}
\begin{center}
\includegraphics[width=0.8\hsize,height=0.5\hsize]{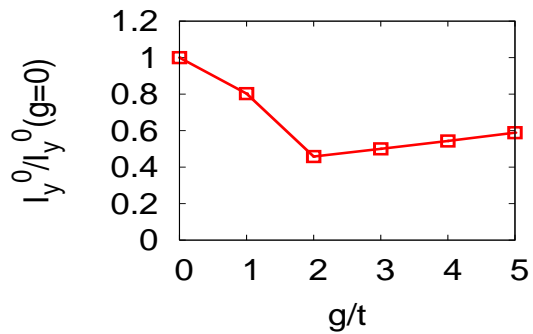}
\caption{The surface current 
$I_y^0$ normalized by $I_y^0(g=0)$ in the absence of
surface disorder, $V_0=0$.
}
\label{fig:Ig_AHsf}
\end{center}
\end{figure}

We show the surface current as a function of $g$ at $V_0=0$ in
Fig.~\ref{fig:Ig_AHsf}.
The surface current is suppressed by the $p_y$-wave superfluidity.
It is noted that similar behaviors are also seen for spatially
uniform gap functions whose amplitudes are chosen to be consistent with
the self consistent calculations.
For non-self-consistent gap functions, 
we can tune the gap amplitudes $\Delta_{1,2}$
for each orbital independently in order to
investigate $\Delta_{1,2}$-dependence of the surface current.
By calculating the surface currents for such $\Delta_{1,2}$ 
we see that
$I_y^0$ is determined by detailed balance between the gap
amplitudes for the two orbitals (not shown).
In the self consistent calculations, ratio between $\Delta_{1,2}$
is determined by the gap equation, which then leads to the
non-monotonic behavior of $I_y^0(g)$ (Fig.~\ref{fig:Ig_AHsf}).

\begin{figure}
\begin{center}
\includegraphics[width=0.8\hsize,height=0.5\hsize]{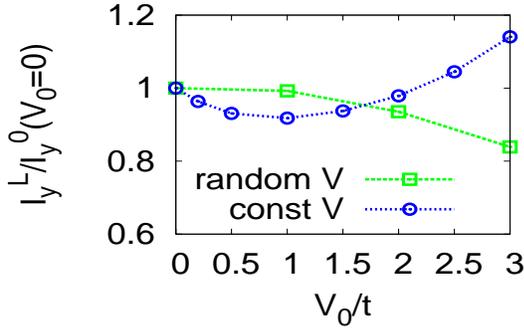}
\caption{The surface current
$I_y^L$ normalized by $I_y^0(V_0=0)$ when $g=5t$.
}
\label{fig:Iv_AHSF}
\end{center}
\end{figure}

\begin{figure}
\begin{center}
\includegraphics[width=0.8\hsize,height=0.5\hsize]{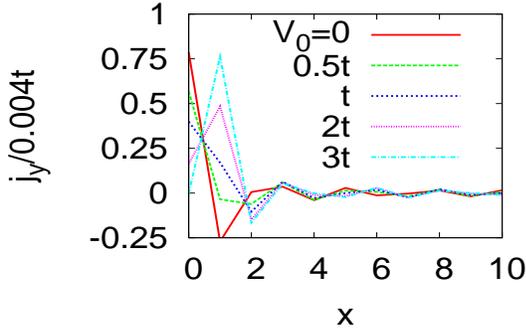}
\caption{The current density $j_y$ near the left surface
for the constant surface potential $(V_{i1},V_{i2})=(V_0,0)$.
}
\label{fig:jy_constV}
\end{center}
\end{figure}

In the presence of the surface perturbation potential $V_{ill'}$,
the current density is modified as in the previous section.
We find that
the resulting left surface current $I_y^L$ can also be changed from $I_y^0$
in contrast to U(1) symmetric systems,
while $I_y^R$ is unchanged.
For the random potential, the surface current $I_y^L$ is suppressed
as shown in Fig.~\ref{fig:Iv_AHSF}.
$I_y^L$ is almost unchanged up tp $V_0\simeq t$, and it decreases
by further increasing $V_0$.
The reduction of $I_y^L$ by the surface roughness well agrees with our
naive expectation that disorder would generally suppress surface currents.
On the other hand, for the constant potential along the left surface sites,
$(V_{i1},V_{i2},V_{i12})=(V_0,0,0)$, 
the surface current $I_y^L$ 
shows non-monotonic $V_0$-dependence.
$I_y^L$ is quickly suppressed as $V_0$ is introduced, 
and then it turns to increase exceeding $I_y^0$ when $V_0$ is 
sufficiently large.
In order to understand this behavior, we show the current density $j_{yi}$
in Fig.~\ref{fig:jy_constV}.
As $V_0$ is increased, $j_{yi}$ gets suppressed at 
the left surface sites $i=(x=0,y)$,
but at the same time, it is increased at the next surface sites $i=(x=1,y)$.
The reduction at $x=0$ sites determines $I_y^L$ for small $V_0\ll t$, while
$I_y^L$ is dominated by the contribution from $x=1$ sites for large $V_0$.
Since we now do not have U(1) symmetry and an associated sum rule,
these changes in $j_y$ do not necessarily cancel out and indeed they add up to
give non-trivial finite values.
Therefore, $I_y^L$ deviates from $I_y^0$ and shows the non-monotonic 
behavior in the present system.


\subsubsection{chiral $p$-wave superfluid}
As a second example of superfluids without time reversal symmetry, 
we study a chiral $p$-wave
superfluid on a square lattice in which surface current is
generated by the superfluidity itself.
The problem of spontaneous surface current and corresponding
OAM, often referred to as ``intrinsic angular momentum paradox", 
has been discussed for more than 40 years
~\cite{pap:Leggett1975,book:Vollhart1990,
book:Volovik2003,book:Leggett2006,pap:Ishikawa1980,pap:Mermin1980,
pap:Kita1998,pap:Goryo1998,
pap:Furusaki2001,pap:Stone2004,pap:Mizushima2008,
pap:Sauls2011,pap:Bradlyn2012,pap:Shitade2014,pap:Hoyos2014,
pap:Tsutsumi2014,pap:Volovik2015,pap:Tada2015,pap:Huang2014}.
Although
most of the previous studies focus only on the weak coupling BCS region,
here we discuss both the BCS region and the BEC region on an equal footing.
In the present study,
similarly to the previous models, surface roughness is introduced as a
typical example of surface perturbations.
We consider the following Hamiltonian
\begin{align}
H_{p\rm SF}&=\sum_{\rm NN} -t_{ij}c^{\dagger}_{i\sigma}c_{j\sigma}
-g\sum_{{\rm NN}}n_{i\sigma}n_{j\bar{\sigma}}+H_{\rm surf},\\
H_{\rm surf}&=\sum V_in_{i\sigma}\notag,
\end{align}
where $V_{i}$ is finite only at the left surface, $i=(x=1,y)$.
The hopping and interaction are allowed only for the nearest neighbor sites,
and the same cylinder geometry as in the previous sections
is used.
The system size is $N_x\times N_y=60\times 20$, and we have confirmed
finite size effects are negligibly small for this size by performing
similar calculations for other sizes. 
We perform mean field calculations of the superfluidity, which
is a good approximation even for a large $g$ 
at zero temperature since there are no
thermal fluctuations in the ground states
~\cite{pap:Eagles1969,book:Leggett1980,pap:BCSBECreview2008}.
The current density is simply
\begin{align}
j_{\mu i}&=\sum_{\sigma}
-it[c^{\dagger}_{i\sigma}c_{i+\hat{\mu}\sigma}
-c^{\dagger}_{i+\hat{\mu}\sigma}c_{i\sigma}].
\end{align}

\begin{figure}
\begin{center}
\includegraphics[width=0.8\hsize,height=0.6\hsize]{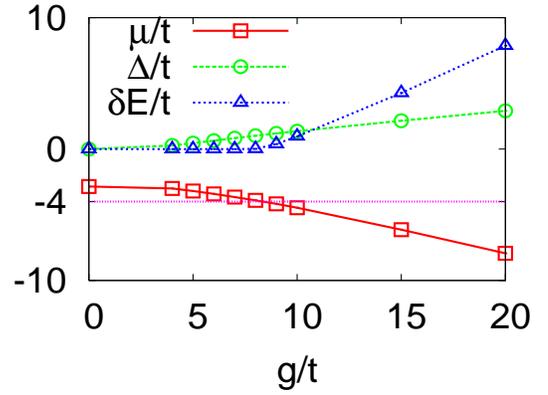}
\caption{
The chemical potential $\mu$, amplitude of the gap functions
in the bulk $\Delta$, 
and spectrum gap $\delta E$ for
filling $n=0.2$.
The purple dotted line represent $\mu_c=-4t$.
}
\label{fig:g}
\end{center}
\end{figure}

Before discussing surface disorder effects,
we first examine basic properties of the model in the absence of
surface disorder.
The present model exhibits a quantum phase transition
~\cite{pap:Read2000,pap:Massignan2010}
when $\mu=\mu_c=-4t$.
For $|\mu|<|\mu_c|$, 
the system is in the BCS region where there are gapless chiral edge modes
at the surfaces,
while for $|\mu|>|\mu_c|$,
the system is 
in the BEC region where there is a spectrum gap.
In Fig.~\ref{fig:g}, we show
the chemical potential $\mu$, amplitudes of the gap functions
at a center site of the system $\Delta$, 
and spectrum gap $\delta E$ for
fixed filling $n=0.2$.
The quantum phase transition takes place around 
$g\simeq 8t-9t$
for this filling where $\mu$ crosses $\mu_c$, and we have confirmed that 
similar behaviors are seen for other low filling.
It is noted that, when filling is high, $n\simeq 1$, the ground state
stays in the BCS region even for large $g$ and
it is hard to realize the
BCS-BEC phase transition.

\begin{figure}
\begin{center}
\includegraphics[width=0.8\hsize,height=0.5\hsize]{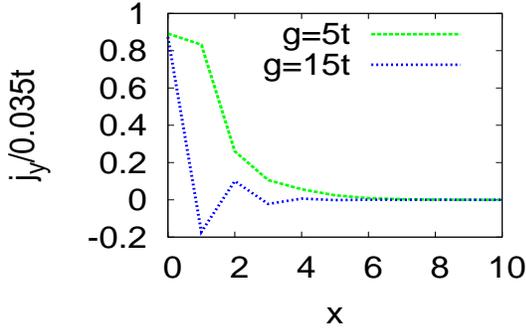}
\caption{
The current density $j_{y}$ near the left surface for $V_0=0$ at $n=0.2$
with $g=5t$ (BCS state, green) and $g=15t$ (BEC region, blue).
}
\label{fig:jsf0}
\end{center}
\end{figure}

\begin{figure}
\begin{center}
\includegraphics[width=0.8\hsize,height=0.5\hsize]{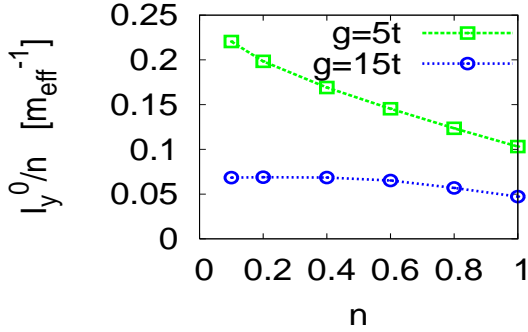}
\caption{
The surface current for different filling $n$ at $g=5t$ and $g=15t$
in unit of the effective mass $m_{\rm eff}=1/(2ta^2)$.
}
\label{fig:In}
\end{center}
\end{figure}

We show the current density $j_{y}$ at $n=0.2$
for $g=5t$ (BCS region) and $g=15t$ (BEC region) in Fig.~\ref{fig:jsf0}
as an example. $j_x$ vanishes everywhere in the system.
The current density in the BEC region 
is more strongly localized near the surface
than that in the BCS region, and it oscillates in sign 
depending on the distance from the surface.
As a result, the surface current in the BEC region is smaller 
than that in the BCS region in the present lattice model,
as shown in Fig.~\ref{fig:In}.
We see that
the overall behavior of $I_y^0$ in the BCS region
as a function of filling $n$
is consistent with the recent work~\cite{com:Tsuruta}.
The surface current for low filling and weak coupling limit approaches
$I=n/(4m_{\rm eff})$ where
$m_{\rm eff}$ is an effective mass $m_{\rm eff}=1/(2ta^2)$ 
with the lattice constant $a$. 
Under an assumption that the surface current is constant along a
boundary of a finite system,
this gives the OAM $L_z=m_{\rm eff}\oint [\vecc{r}\times \vecc{I}]_zdl
=N/2$ where $N$ is the total
number of fermions in agreement with the previous calculations
for continuum systems without lattice potentials
~\cite{pap:Sauls2011,pap:Tsutsumi2014,pap:Volovik2015,
pap:Tada2015}.
In the present lattice model,
in contrast to the continuum systems where $L_z=N/2$ holds both
in the weak coupling region and strong coupling region, 
$I_y$ and the corresponding OAM $L_z$ is decreased when the coupling 
constant $g$ is increased. 
This is because, even for low filling $n\ll 1$ where 
lattice effects is expected to be less important,
the smallest size of a bosonic molecule of two fermions
is bounded by the lattice constant
for a non-$s$-wave superfluid on a lattice,
and therefore, presence of a lattice is significant especially for
the BEC region rather than the BCS region.
Because of this lattice effect, 
the surface current per fermion for strong $g=15t$
is almost independent of filling as seen in Fig.~\ref{fig:In},
although the system stays in the BCS region for high filling
$n \simeq 1$.

\begin{figure}
\begin{center}
\includegraphics[width=0.8\hsize,height=0.5\hsize]{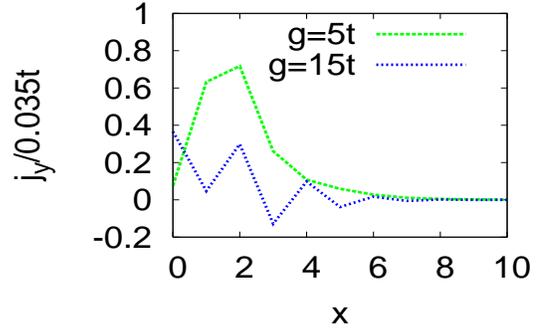}
\caption{
The current density $j_{y}$ near the left surface for $V_0=8t$ at $n=0.2$
with $g=5t$ (BCS state, green) and $g=15t$ (BEC region, blue).
}
\label{fig:jsfav}
\end{center}
\end{figure}

Now, we discuss effects of the surface disorder.
The current density $j_{y}$ averaged over 
disorder configurations for $n=0.2$ at a large $V_0=8t$
is shown in Fig.~\ref{fig:jsfav}. 
For the BCS region, $j_y$ is suppressed especially at the surface sites $x=0$
and it is enhanced at inner sites $x=3,4$ to partly compensate the reduction,
while $j_y$ in the BEC region becomes strongly oscillating and
effects  of $V_{i}$ propagate into further inner sites $x=4,5$.
As in the previous model \eqref{eq:Hahs}, change of $j_y$ does not need to obey
a sum rule and the surface current can be modified from $I_y^0$.
\begin{figure}
\begin{center}
\includegraphics[width=0.8\hsize,height=0.5\hsize]{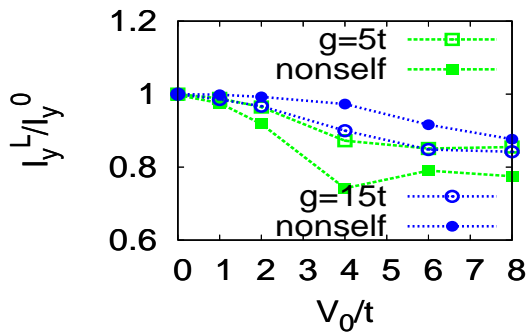}
\caption{
The left surface current $I_y^L$ for different $V_0$ at $g=5t$ 
(green square) 
and $g=15t$ (blue circle)
when filling is $n=0.2$.
Filled symbols correspond to 
non-self-consistent gap functions whose amplitudes are
chosen to be consistent with self consistent calculations at $g=5t$
and $g=15t$, respectively.
}
\label{fig:Iv}
\end{center}
\end{figure}
Indeed, as shown in Fig.~\ref{fig:Iv},
the surface current is suppressed by the surface disorder.
The decrease of $I_y^L$ in the weak coupling BCS limit is consistent
with the previous studies~\cite{pap:Sauls2011,pap:Ashby2009,pap:Nagato2011}.
Interestingly, $I_y^L$ decreases not only in the BCS region
but also in the BEC region.
This would be because the smallest bosonic molecule size is bounded by the
lattice constant and the disorder length scale is of the same order
in the present model.
We have confirmed similar behaviors of $I_y^L$ for 
different parameter sets $(g,n)$.
For a comparison, we also calculate surface currents
for non-self-consistent gap functions 
which are constant in space and whose amplitudes are chosen to be
consistent with the self-consistent calculations.
The surface currents are suppressed in a similar way as in 
the self consistent calculations, 
which means that change of the gap functions 
around the surface
by $H_{\rm surf}$ is not important for the reduction of $I_y$.
We note that, although the reduction of $I_y^L$ by the surface
disorder is moderate,
it was pointed out that,
for domain boundaries with opposite chiralities in the BCS region, 
the boundary current strongly relies on boundary conditions and it
can change
even its direction~\cite{pap:Tsutsumi2014,pap:Volovik2015}.

\subsection{discussion}
We have discussed suppression of surface currents in cylinder systems
in the previous sections.
For a realistic finite size 
system with open boundary conditions for all directions,
surface current flows along the surface and generates global rotation.
In the absence of surface perturbations, its magnitude is obviously uniform
along the surface.
If some parts of the surface are perturbed,
the surface current would be changed not only at the perturbed surface
but also at the whole surface,
because of the continuity of the current density.
Therefore, associated OAM would also be modified.

Although we have examined particular realizations of surface potentials
among possible surface perturbations, 
existence of surface perturbations which change $I_y$ 
conceptually distinguishes a system without U(1) symmetry from 
a system with U(1) symmetry.
The surface current is not uniquely
determined as a bulk property and
there may exist surface 
perturbations which drastically change it
in the former,
while the surface current is an intrinsic quantity in the latter.
The absence of a sum rule for the surface current density 
and the numerical results 
suggest that there is no bulk-surface correspondence 
for surface currents and corresponding OAM in superfluids, 
and surface conditions should be fully taken into account 
in order to calculate these quantities
in contrast to some of the previous studies for chiral superfluids
~\cite{pap:Bradlyn2012,pap:Shitade2014}.
From these discussions, it is considered that
surface currents and OAM
in superfluids with broken time reversal symmetry would be 
subtle quantities, and
experimentally, one needs to control surface conditions carefully
in order to measure these quantities.

Let us briefly discuss the change of surface currents and OAM
in superfluids in view of thermodynamics.
In the present study, we have
implicitly assumed that
the lattices (or containers for non-lattice systems) are at rest
in the laboratory frame
by some reasons.
If the lattice or container is fixed spatially to a much larger
environment, the system composed only of 
the superfluid and lattice does not conserve
the OAM and it is an open system with respect to angular momentum. 
Although
OAM of the superfluid alone can be changed by surface perturbations,
total OAM of the whole system including the 
large environment should be a conserved 
thermodynamic quantity and is independent of surface details 
of the lattice/container.
If the system composed of a superfliud and a lattice/container 
is suspended in the midair
and set to be at rest, 
as temperature is decreased 
down to the superfluid transition temperature,
the lattice/container should start to rotate in an opposite direction to the
superfluid rotation
in order to keep the total OAM
of the whole system zero due to the angular momentum conservation.
This is analogous to the Einstein-de Haas effect
and each of the OAM for the superfluid and lattice/container
would depend on surface conditions in the present system.

In the present study, we have not taken into account
the electromagnetic field which couples to charged particles.
Indeed,
it is especially important in superconductors which exhibit
Meissner effect.
If Meissner effect is included, 
current density distributions are modified and
net surface currents would vanish for uniform superconducting states
\cite{pap:Furusaki2001,pap:Ashby2009}.
The problem of Meissner effect would be more complicated 
in a system where there exist circulating currents even in 
non-superconducting states, such as the model \eqref{eq:Hahs}
and ferromagnetic superconductors.
This issue is left for a future work.

When we were finalizing the present paper,
we became aware of a relevant article by ~\textcite{pap:Kusama1999}
which claims that, when surface currents are not canceled between
left and right surfaces in a cylinder, supercurrent will compensate this and 
total current will vanish in superfluids.
Although this is an important possibility for a vanishing total current,
this issue has not been well understood.
For example, in \textcite{pap:Kusama1999},
the current density even near the non-perturbed surface is
strongly changed from the original configuration if supercurrent is included.
This seems unphysical, 
since effects of local perturbations only on a surface should not propagate to
the opposite surface.
Secondly, although 
they compared the free energy density of different system sizes
for a technical reason, this cannot be justified in general.
Their discussion is motivated by the original Bloch-Bohm's theorem which
is not helpful for surface currents as shown in Sec.~\ref{sec:BlochBohm},
and further investigations would be required to understand
this issue.
Here, in order to have an insight, 
let us breify consider a possible supercurrent in a cylinder 
$L\times L$ where open (periodic) boundary condition is 
imposed for $x(y)$-direction, and surface perturbations are
introduced only for one surface.
Supercurrent density is roughly proportional to 
$\nabla \phi(x,y)$ for a gap function
$\Delta=e^{i\phi}\tilde{\Delta}$ where $\tilde{\Delta}$ is a gap
function without a modulation.
In the cylinder considered, the supercurrent density $\sim \nabla\phi$ should
be translationally symmetric along the $y$-direction.
Therefore, $\phi$ must be $\phi=q_xx+q_yy$ with 
$q_{\mu}=2\pi n/L, (n=0,\pm1,\cdots)$,
and $q_x$ should be zero for a vanishing 
total supercurrent in the $x$-direction.
Besides, 
$q_y$ should be of order $2\pi/L$ so that
the current density near the non-perturbed surface remains unchanged.
In such a case, $q_y\simeq 2\pi/L$ introduces supercurrent density
$\sim 1/L$ at every site, resulting in a total supercurrent $\sim L$ 
($L^{d-1}$ in $d$-dimensions) in the whole system, which is the same order as
the surface current.
We note that,
however, two thermodynamic states constructed from wavefunctions with 
$\Delta$ and $\tilde{\Delta}$ respectively
could not be distinguished by local operators and therefore
they converge to a single state as in Sec. 
\ref{sec:BlochBohm}.
Besides,
if we consider a semi-infinite system and discuss it within 
weak coupling approximations
as in the previous studies~\cite{pap:Ashby2009,pap:Nagato2011}, 
we could not impose a boundary condition at infinite
with an infinitesimal supercurrent density.
Furthermore, if we consider a realistic finite size sample with boundaries
and introduce surface roughness, 
it is impossible to realize a uniform supercurrent density and 
possible current density
configuration especially near the surface would be quite complicated.
Therefore, possible compensation by supercurrent is a subtle issue.
In order to discuss such a subtle issue,
we would 
also have to be careful about validity of the mean field approximations.
Further investigations of gap functions and supercurrent may be required 
for clarifying a role of supercurrents.

\section{summary}
\label{sec:summary}
In summary, we have investigated equilibrium
surface currents in systems 
with or without U(1) particle number conservation.
For the systems with U(1) symmetry,
we showed 
that the surface currents are
independent of surface perturbations
based on the sum rule for current densities, which was confirmed by
numerical calculations for a concrete model with surface
perturbations.
Therefore, 
the surface currents and corresponding orbital 
magnetization are bulk quantities which are robust against
surface conditions.
The sum rule argument is also applicable to the Landau diamagnetism,
and it would give a new understanding on
the known equivalence between the bulk approaches and
the surface approaches.
On the other hand, in superfluids which do not
have U(1) symmetry,
the surface currents are changed by surface perturbations.
Especially, in a chiral superfluid on a lattice,
the surface current is suppressed 
by surface disorder
not only in the weak coupling BCS region but also
in the strong coupling BEC region.
These results imply that surface mass currents and orbital angular momentum
in superfluids with broken time reversal symmetry would be subtle quantities
and depend on surface details.
Experimentally, one needs to control surface conditions carefully
in order to measure these quantities.

\begin{acknowledgements}
We thank M. Oshikawa, H. Akai,
S. Fujimoto, Y. Yanase, T. Osada, S. Sugiura, Y. Nishida,
M. Sigrist, and A. H. MacDonald for valuable discussions.
This work was partly supported by Grant-in-Aid for Scientific Research 
(Nos. 26800177, 25103706) and by a Grant-in-Aid for 
Program for Advancing Strategic International Networks to Accelerate the Circulation of Talented Researchers (No. R2604) ``TopoNet".
\end{acknowledgements}

\appendix
\subsection*{Appendix A. Sum Rule for Continuum Model}
\renewcommand{\theequation}{A\arabic{equation}}
\setcounter{equation}{0}
Our sum rule argument holds also for continuum models with
lattice potentials.
We consider a general Hamiltonian defined on a cylinder $L\times L$
with the open (periodic) boundary condition for $x(y)$-direction,
\begin{align}
H&=\int_{L^2} d^2x \psi^{\dagger}{\mathcal K}\psi+H_{\rm int}+H_{\rm surf},
\end{align}
where ${\mathcal K}$ is the single-particle Hamiltonian
including the lattice potential
and $\psi(\vecc{r})=(\psi_{\uparrow}(\vecc{r}),\psi_{\downarrow}
(\vecc{r}))$ is the fermionic field operator.
Then we expand the field operator in terms of non-interacting 
single-particle wavefunctions,
\begin{align}
&\psi(\vecc{r})=\sum_{k_yn} \varphi_{k_yn}(\vecc{r})c_{k_yn},\\
&{\mathcal K}\varphi_{k_yn}=\varepsilon_{k_yn}\varphi_{k_yn},
\end{align}
where $k_y$ is Bloch wavenumber along the $y$-direction 
and $n$ represents other
indices including a quantum number corresponding to 
position $x$.
The Matsubara Green's function
$G(r,r')=-\langle\langle \psi(r)\psi^{\dagger}(r')\rangle\rangle$
is also expanded as
\begin{align}
G(\vecc{r},\vecc{r}',i\omega)&=\sum_{k_ynn'}\varphi_{k_yn}(\vecc{r})
g_{nn'}(k_y,i\omega)\varphi_{k_yn'}^{\ast}(\vecc{r}'),\\
\hat{g}(k_y,i\omega,)
&=[\hat{g}_0^{-1}-\hat{\Sigma}]^{-1},
\end{align}
where $(g_0)^{-1}_{nn'}=[i\omega-\varepsilon_{k_yn}]\delta_{nn'}$
is the non-interacting Green's function and
$\hat{\Sigma}$ is the selfenergy in the $\varphi_{k_yn}$-basis.
Similarly to the lattice models, spontaneous symmetry breaking 
order parameters are easily incorporated into $g$.
In the cylinder,
the surface current averaged over the $y$-direction
is simply given by 
\begin{align}
I_y^{L(R)}&=\frac{1}{L}\int_{S_L(S_R)}d^2xj_{y}(\vecc{r})\notag\\
&=\frac{1}{L}\sum_{i\in S_L(S_R)}\int_{v_i}d^2xj_y(\vecc{r}),
\label{eqA:I}
\end{align}
where $j_y$ is the current density.
As discussed in the main text and Ref.~\onlinecite{com:j},
contributions to the surface current come only from $S_{L}(S_R)$
and not from bulk regions.
Therefore, when U(1) charge symmetry is present,
the total surface current $I_y^{\rm tot}=\langle I_y^L+I_y^R\rangle$
is written as
\begin{align}
I_y^{\rm tot}&=
\frac{1}{L}\int_{L^2}d^2x\langle j_{y}(\vecc{r})\rangle \notag\\
&=\frac{1}{L}\int_{L^2}d^2x\frac{-i}{2m}[\partial_{y'}-\partial_{y}]
G(\vecc{r},\vecc{r}',\tau=0^-)|_{r=r'}\notag\\
&=-\frac{1}{L}\sum_{k_y}{\rm tr}\Bigl[
\frac{\partial \hat{g}_0^{-1}(k)}{\partial k_y} \hat{g}(k)
\Bigr].
\end{align}
Here, trace describes summation over $n$ and
$\omega$.
We can now follow the same argument as in the main text,
and show $I^{\rm tot}_y=0$ even in the presence of left surface
perturbations.
This means that the surface current is unchanged by 
surface perturbations.

It is noted that the surface currents in lattice models are 
obtained by the following replacement in Eq.~\eqref{eqA:I},
\begin{align}
&\sum_{i\in {\rm surf}}\int_{v_i}d^2x\psi^{\dagger}(-i\partial_y)
\psi\notag\\
&=\sum_{i\in {\rm surf}}\sum_{RR',ll'}\langle w_{Rl}|-i\partial_y
|w_{R'l'}\rangle_{v_i}c^{\dagger}_{Rl}c_{R'l'}\notag\\
&\rightarrow \sum_{RR'\in {\rm surf}}\sum_{ll'}\langle w_{Rl}|-i\partial_y
|w_{R'l'}\rangle c^{\dagger}_{Rl}c_{R'l'},
\end{align}
where $w_{Rl}$ is a Wannier function and 
$\langle\cdots\rangle_{v_i}=\int_{v_i} d^2x$.
This replacement is verified when the chosen Wannier function
is well localized in a length scale which is much smaller than 
the system size. 
By similar replacements, other local site quantities
such as particle density at site $i$ become equivalent to
the usual Wannier basis descriptions,
$\int_{v_i}d^2x\psi^{\dagger}\psi=\sum_{RR'll'}
\langle w_{Rl}|w_{R'l'}\rangle_{v_i}c^{\dagger}_{Rl}c_{R'l'}
\rightarrow \sum_{R=R'\in v_i}\sum_{ll'}
\langle w_{Rl}|w_{R'l'}\rangle c^{\dagger}_{Rl}c_{R'l'}
=\sum_{l}c^{\dagger}_{R_il}c_{R_il}$.
Even when the replacement of the surface current operator 
is legitimate, however, the resulting 
surface current might depend on Wannier functions or gauge of 
Bloch functions~\cite{pap:Thonhauser2005,pap:Ceresoli2006},
although the original definition~\eqref{eqA:I} is 
independent of them.

\bibliography{surf}

\begin{thebibliography}{76}
\expandafter\ifx\csname natexlab\endcsname\relax\def\natexlab#1{#1}\fi
\expandafter\ifx\csname bibnamefont\endcsname\relax
  \def\bibnamefont#1{#1}\fi
\expandafter\ifx\csname bibfnamefont\endcsname\relax
  \def\bibfnamefont#1{#1}\fi
\expandafter\ifx\csname citenamefont\endcsname\relax
  \def\citenamefont#1{#1}\fi
\expandafter\ifx\csname url\endcsname\relax
  \def\url#1{\texttt{#1}}\fi
\expandafter\ifx\csname urlprefix\endcsname\relax\def\urlprefix{URL }\fi
\providecommand{\bibinfo}[2]{#2}
\providecommand{\eprint}[2][]{\url{#2}}

\bibitem[{\citenamefont{Martin}(2004)}]{book:Martin2004}
\bibinfo{author}{\bibfnamefont{R.~M.} \bibnamefont{Martin}},
  \emph{\bibinfo{title}{Electronic Structure: Basic Theory and Practical
  Methods}} (\bibinfo{publisher}{Cambridge University Press},
  \bibinfo{year}{2004}), \bibinfo{edition}{1st} ed.

\bibitem[{\citenamefont{Hasan and Kane}(2010)}]{pap:Hasan2010}
\bibinfo{author}{\bibfnamefont{M.~Z.} \bibnamefont{Hasan}} \bibnamefont{and}
  \bibinfo{author}{\bibfnamefont{C.~L.} \bibnamefont{Kane}},
  \bibinfo{journal}{Rev.~Mod.~Phys.} \textbf{\bibinfo{volume}{82}},
  \bibinfo{pages}{3045} (\bibinfo{year}{2010}).

\bibitem[{\citenamefont{Qi and Zhang}(2011)}]{pap:Qi2011}
\bibinfo{author}{\bibfnamefont{X.~L.} \bibnamefont{Qi}} \bibnamefont{and}
  \bibinfo{author}{\bibfnamefont{S.~C.} \bibnamefont{Zhang}},
  \bibinfo{journal}{Rev.~Mod.~Phys.} \textbf{\bibinfo{volume}{83}},
  \bibinfo{pages}{1057} (\bibinfo{year}{2011}).

\bibitem[{\citenamefont{Wen}(2004)}]{book:Wen2004}
\bibinfo{author}{\bibfnamefont{X.~G.} \bibnamefont{Wen}},
  \emph{\bibinfo{title}{Quantum Field Theory of Many-Body Systems}}
  (\bibinfo{publisher}{Oxford University Press}, \bibinfo{year}{2004}),
  \bibinfo{edition}{1st} ed.

\bibitem[{\citenamefont{Gat and Avron}(2003{\natexlab{a}})}]{pap:Gat2003a}
\bibinfo{author}{\bibfnamefont{O.}~\bibnamefont{Gat}} \bibnamefont{and}
  \bibinfo{author}{\bibfnamefont{J.~E.} \bibnamefont{Avron}},
  \bibinfo{journal}{Phys.~Rev.~Lett.} \textbf{\bibinfo{volume}{91}},
  \bibinfo{pages}{186801} (\bibinfo{year}{2003}{\natexlab{a}}).

\bibitem[{\citenamefont{Gat and Avron}(2003{\natexlab{b}})}]{pap:Gat2003b}
\bibinfo{author}{\bibfnamefont{O.}~\bibnamefont{Gat}} \bibnamefont{and}
  \bibinfo{author}{\bibfnamefont{J.~E.} \bibnamefont{Avron}},
  \bibinfo{journal}{New J. Phys.} \textbf{\bibinfo{volume}{5}},
  \bibinfo{pages}{44} (\bibinfo{year}{2003}{\natexlab{b}}).

\bibitem[{\citenamefont{Xiao et~al.}(2005)\citenamefont{Xiao, Shi, and
  Niu}}]{pap:Xiao2005}
\bibinfo{author}{\bibfnamefont{D.}~\bibnamefont{Xiao}},
  \bibinfo{author}{\bibfnamefont{J.}~\bibnamefont{Shi}}, \bibnamefont{and}
  \bibinfo{author}{\bibfnamefont{Q.}~\bibnamefont{Niu}},
  \bibinfo{journal}{Phys.~Rev.~Lett.} \textbf{\bibinfo{volume}{95}},
  \bibinfo{pages}{137204} (\bibinfo{year}{2005}).

\bibitem[{\citenamefont{Thonhauser et~al.}(2005)\citenamefont{Thonhauser,
  Ceresoli, Vanderbilt, and Resta}}]{pap:Thonhauser2005}
\bibinfo{author}{\bibfnamefont{T.}~\bibnamefont{Thonhauser}},
  \bibinfo{author}{\bibfnamefont{D.}~\bibnamefont{Ceresoli}},
  \bibinfo{author}{\bibfnamefont{D.}~\bibnamefont{Vanderbilt}},
  \bibnamefont{and} \bibinfo{author}{\bibfnamefont{R.}~\bibnamefont{Resta}},
  \bibinfo{journal}{Phys.~Rev.~Lett.} \textbf{\bibinfo{volume}{95}},
  \bibinfo{pages}{137205} (\bibinfo{year}{2005}).

\bibitem[{\citenamefont{Ceresoli et~al.}(2006)\citenamefont{Ceresoli,
  Thonhauser, Vanderbilt, and Resta}}]{pap:Ceresoli2006}
\bibinfo{author}{\bibfnamefont{D.}~\bibnamefont{Ceresoli}},
  \bibinfo{author}{\bibfnamefont{T.}~\bibnamefont{Thonhauser}},
  \bibinfo{author}{\bibfnamefont{D.}~\bibnamefont{Vanderbilt}},
  \bibnamefont{and} \bibinfo{author}{\bibfnamefont{R.}~\bibnamefont{Resta}},
  \bibinfo{journal}{Phys.~Rev.~B} \textbf{\bibinfo{volume}{74}},
  \bibinfo{pages}{024408} (\bibinfo{year}{2006}).

\bibitem[{\citenamefont{Shi et~al.}(2007)\citenamefont{Shi, Vignale, Xiao, and
  Niu}}]{pap:Shi2007}
\bibinfo{author}{\bibfnamefont{J.}~\bibnamefont{Shi}},
  \bibinfo{author}{\bibfnamefont{G.}~\bibnamefont{Vignale}},
  \bibinfo{author}{\bibfnamefont{D.}~\bibnamefont{Xiao}}, \bibnamefont{and}
  \bibinfo{author}{\bibfnamefont{Q.}~\bibnamefont{Niu}},
  \bibinfo{journal}{Phys.~Rev.~Lett.} \textbf{\bibinfo{volume}{99}},
  \bibinfo{pages}{197202} (\bibinfo{year}{2007}).

\bibitem[{\citenamefont{Resta}(2010)}]{pap:Resta2010}
\bibinfo{author}{\bibfnamefont{R.}~\bibnamefont{Resta}}, \bibinfo{journal}{J.
  Phys.:Condens. Matter} \textbf{\bibinfo{volume}{22}}, \bibinfo{pages}{123201}
  (\bibinfo{year}{2010}).

\bibitem[{\citenamefont{Thonhauser}(2011)}]{pap:Thonhauser2011}
\bibinfo{author}{\bibfnamefont{T.}~\bibnamefont{Thonhauser}},
  \bibinfo{journal}{Int. J. Mod. Phys. B} \textbf{\bibinfo{volume}{25}},
  \bibinfo{pages}{1429} (\bibinfo{year}{2011}).

\bibitem[{\citenamefont{Xiao et~al.}(2010)\citenamefont{Xiao, Chang, and
  Niu}}]{pap:Xiao2010}
\bibinfo{author}{\bibfnamefont{D.}~\bibnamefont{Xiao}},
  \bibinfo{author}{\bibfnamefont{M.~C.} \bibnamefont{Chang}}, \bibnamefont{and}
  \bibinfo{author}{\bibfnamefont{Q.}~\bibnamefont{Niu}},
  \bibinfo{journal}{Rev.~Mod.~Phys.} \textbf{\bibinfo{volume}{82}},
  \bibinfo{pages}{1959} (\bibinfo{year}{2010}).

\bibitem[{\citenamefont{Matsumoto and Murakami}(2011)}]{pap:Matsumoto2011}
\bibinfo{author}{\bibfnamefont{R.}~\bibnamefont{Matsumoto}} \bibnamefont{and}
  \bibinfo{author}{\bibfnamefont{S.}~\bibnamefont{Murakami}},
  \bibinfo{journal}{Phys.~Rev.~Lett.} \textbf{\bibinfo{volume}{106}},
  \bibinfo{pages}{197202} (\bibinfo{year}{2011}).

\bibitem[{\citenamefont{Chen and Lee}(2011)}]{pap:Chen2011}
\bibinfo{author}{\bibfnamefont{K.~T.} \bibnamefont{Chen}} \bibnamefont{and}
  \bibinfo{author}{\bibfnamefont{P.~A.} \bibnamefont{Lee}},
  \bibinfo{journal}{Phys.~Rev.~B} \textbf{\bibinfo{volume}{84}},
  \bibinfo{pages}{205137} (\bibinfo{year}{2011}).

\bibitem[{\citenamefont{Chen and Lee}(2012)}]{pap:Chen2012}
\bibinfo{author}{\bibfnamefont{K.~T.} \bibnamefont{Chen}} \bibnamefont{and}
  \bibinfo{author}{\bibfnamefont{P.~A.} \bibnamefont{Lee}},
  \bibinfo{journal}{Phys.~Rev.~B} \textbf{\bibinfo{volume}{86}},
  \bibinfo{pages}{195111} (\bibinfo{year}{2012}).

\bibitem[{\citenamefont{Leggett}(1975)}]{pap:Leggett1975}
\bibinfo{author}{\bibfnamefont{A.~J.} \bibnamefont{Leggett}},
  \bibinfo{journal}{Rev.~Mod.~Phys.} \textbf{\bibinfo{volume}{47}},
  \bibinfo{pages}{331} (\bibinfo{year}{1975}).

\bibitem[{\citenamefont{Vollhart and W{\"o}lfle}(1990)}]{book:Vollhart1990}
\bibinfo{author}{\bibfnamefont{D.}~\bibnamefont{Vollhart}} \bibnamefont{and}
  \bibinfo{author}{\bibfnamefont{P.}~\bibnamefont{W{\"o}lfle}},
  \emph{\bibinfo{title}{The Superfluid Phase of Helium 3}}
  (\bibinfo{publisher}{Taylor and Francis, London}, \bibinfo{year}{1990}),
  \bibinfo{edition}{1st} ed.

\bibitem[{\citenamefont{Volovik}(2003)}]{book:Volovik2003}
\bibinfo{author}{\bibfnamefont{G.~E.} \bibnamefont{Volovik}},
  \emph{\bibinfo{title}{The Universe in a Helium Droplet}}
  (\bibinfo{publisher}{Oxford University Press, Oxford}, \bibinfo{year}{2003}).

\bibitem[{\citenamefont{Leggett}(2006)}]{book:Leggett2006}
\bibinfo{author}{\bibfnamefont{A.~J.} \bibnamefont{Leggett}},
  \emph{\bibinfo{title}{Quantum Liquids: Bose Condensation and Cooper Pairing
  in Condensed-matter Systems}} (\bibinfo{publisher}{Oxford University Press,
  Oxford}, \bibinfo{year}{2006}).

\bibitem[{\citenamefont{Ishikawa et~al.}(1980)\citenamefont{Ishikawa, Miyake,
  and Usui}}]{pap:Ishikawa1980}
\bibinfo{author}{\bibfnamefont{M.}~\bibnamefont{Ishikawa}},
  \bibinfo{author}{\bibfnamefont{K.}~\bibnamefont{Miyake}}, \bibnamefont{and}
  \bibinfo{author}{\bibfnamefont{T.}~\bibnamefont{Usui}},
  \bibinfo{journal}{Prog.~Theor.~Phys.} \textbf{\bibinfo{volume}{63}},
  \bibinfo{pages}{1083} (\bibinfo{year}{1980}).

\bibitem[{\citenamefont{Mermin and Muzikar}(1980)}]{pap:Mermin1980}
\bibinfo{author}{\bibfnamefont{N.~D.} \bibnamefont{Mermin}} \bibnamefont{and}
  \bibinfo{author}{\bibfnamefont{P.}~\bibnamefont{Muzikar}},
  \bibinfo{journal}{Phys.~Rev.~B} \textbf{\bibinfo{volume}{21}},
  \bibinfo{pages}{980} (\bibinfo{year}{1980}).

\bibitem[{\citenamefont{Kita}(1998)}]{pap:Kita1998}
\bibinfo{author}{\bibfnamefont{T.}~\bibnamefont{Kita}},
  \bibinfo{journal}{J.~Phys.~Soc.~Jpn.} \textbf{\bibinfo{volume}{67}},
  \bibinfo{pages}{216} (\bibinfo{year}{1998}).

\bibitem[{\citenamefont{Goryo}(1998)}]{pap:Goryo1998}
\bibinfo{author}{\bibfnamefont{J.}~\bibnamefont{Goryo}},
  \bibinfo{journal}{Phys. Lett. A} \textbf{\bibinfo{volume}{246}},
  \bibinfo{pages}{549} (\bibinfo{year}{1998}).

\bibitem[{\citenamefont{Furusaki et~al.}(2001)\citenamefont{Furusaki,
  Matsumoto, and Sigrist}}]{pap:Furusaki2001}
\bibinfo{author}{\bibfnamefont{A.}~\bibnamefont{Furusaki}},
  \bibinfo{author}{\bibfnamefont{M.}~\bibnamefont{Matsumoto}},
  \bibnamefont{and} \bibinfo{author}{\bibfnamefont{M.}~\bibnamefont{Sigrist}},
  \bibinfo{journal}{Phys.~Rev.~B} \textbf{\bibinfo{volume}{64}},
  \bibinfo{pages}{054514} (\bibinfo{year}{2001}).

\bibitem[{\citenamefont{Stone and Roy}(2004)}]{pap:Stone2004}
\bibinfo{author}{\bibfnamefont{M.}~\bibnamefont{Stone}} \bibnamefont{and}
  \bibinfo{author}{\bibfnamefont{R.}~\bibnamefont{Roy}},
  \bibinfo{journal}{Phys.~Rev.~B} \textbf{\bibinfo{volume}{69}},
  \bibinfo{pages}{184511} (\bibinfo{year}{2004}).

\bibitem[{\citenamefont{Mizushima et~al.}(2008)\citenamefont{Mizushima,
  Ichioka, and Machida}}]{pap:Mizushima2008}
\bibinfo{author}{\bibfnamefont{T.}~\bibnamefont{Mizushima}},
  \bibinfo{author}{\bibfnamefont{M.}~\bibnamefont{Ichioka}}, \bibnamefont{and}
  \bibinfo{author}{\bibfnamefont{K.}~\bibnamefont{Machida}},
  \bibinfo{journal}{Phys.~Rev.~Lett.} \textbf{\bibinfo{volume}{101}},
  \bibinfo{pages}{150409} (\bibinfo{year}{2008}).

\bibitem[{\citenamefont{Sauls}(2011)}]{pap:Sauls2011}
\bibinfo{author}{\bibfnamefont{J.~A.} \bibnamefont{Sauls}},
  \bibinfo{journal}{Phys.~Rev.~B} \textbf{\bibinfo{volume}{84}},
  \bibinfo{pages}{214509} (\bibinfo{year}{2011}).

\bibitem[{\citenamefont{Bradlyn et~al.}(2012)\citenamefont{Bradlyn, Goldstein,
  and Read}}]{pap:Bradlyn2012}
\bibinfo{author}{\bibfnamefont{B.}~\bibnamefont{Bradlyn}},
  \bibinfo{author}{\bibfnamefont{M.}~\bibnamefont{Goldstein}},
  \bibnamefont{and} \bibinfo{author}{\bibfnamefont{N.}~\bibnamefont{Read}},
  \bibinfo{journal}{Phys.~Rev.~B} \textbf{\bibinfo{volume}{86}},
  \bibinfo{pages}{245309} (\bibinfo{year}{2012}).

\bibitem[{\citenamefont{Shitade and Kimura}(2014)}]{pap:Shitade2014}
\bibinfo{author}{\bibfnamefont{A.}~\bibnamefont{Shitade}} \bibnamefont{and}
  \bibinfo{author}{\bibfnamefont{T.}~\bibnamefont{Kimura}},
  \bibinfo{journal}{Phys.~Rev.~B} \textbf{\bibinfo{volume}{90}},
  \bibinfo{pages}{134510} (\bibinfo{year}{2014}).

\bibitem[{\citenamefont{Hoyos et~al.}(2014)\citenamefont{Hoyos, Moroz, and
  Son}}]{pap:Hoyos2014}
\bibinfo{author}{\bibfnamefont{C.}~\bibnamefont{Hoyos}},
  \bibinfo{author}{\bibfnamefont{S.}~\bibnamefont{Moroz}}, \bibnamefont{and}
  \bibinfo{author}{\bibfnamefont{D.~T.} \bibnamefont{Son}},
  \bibinfo{journal}{Phys.~Rev.~B} \textbf{\bibinfo{volume}{89}},
  \bibinfo{pages}{174507} (\bibinfo{year}{2014}).

\bibitem[{\citenamefont{Tsutsumi}(2014)}]{pap:Tsutsumi2014}
\bibinfo{author}{\bibfnamefont{Y.}~\bibnamefont{Tsutsumi}},
  \bibinfo{journal}{J. Low Temp. Phys.} \textbf{\bibinfo{volume}{175}},
  \bibinfo{pages}{51} (\bibinfo{year}{2014}).

\bibitem[{\citenamefont{Volovik}(2015)}]{pap:Volovik2015}
\bibinfo{author}{\bibfnamefont{G.~E.} \bibnamefont{Volovik}},
  \bibinfo{journal}{JETP Letters} \textbf{\bibinfo{volume}{100}},
  \bibinfo{pages}{742} (\bibinfo{year}{2015}).

\bibitem[{\citenamefont{Tada et~al.}(2015)\citenamefont{Tada, Nie, and
  Oshikawa}}]{pap:Tada2015}
\bibinfo{author}{\bibfnamefont{Y.}~\bibnamefont{Tada}},
  \bibinfo{author}{\bibfnamefont{W.}~\bibnamefont{Nie}}, \bibnamefont{and}
  \bibinfo{author}{\bibfnamefont{M.}~\bibnamefont{Oshikawa}},
  \bibinfo{journal}{Phys.~Rev.~Lett.} \textbf{\bibinfo{volume}{114}},
  \bibinfo{pages}{195301} (\bibinfo{year}{2015}).

\bibitem[{\citenamefont{Huang et~al.}(2014)\citenamefont{Huang, Taylor, and
  Kallin}}]{pap:Huang2014}
\bibinfo{author}{\bibfnamefont{W.}~\bibnamefont{Huang}},
  \bibinfo{author}{\bibfnamefont{E.}~\bibnamefont{Taylor}}, \bibnamefont{and}
  \bibinfo{author}{\bibfnamefont{C.}~\bibnamefont{Kallin}},
  \bibinfo{journal}{Phys.~Rev.~B} \textbf{\bibinfo{volume}{90}},
  \bibinfo{pages}{224519} (\bibinfo{year}{2014}).

\bibitem[{\citenamefont{Mackenzie and Maeno}(2003)}]{pap:Maeno2003}
\bibinfo{author}{\bibfnamefont{A.~P.} \bibnamefont{Mackenzie}}
  \bibnamefont{and} \bibinfo{author}{\bibfnamefont{Y.}~\bibnamefont{Maeno}},
  \bibinfo{journal}{Rev.~Mod.~Phys.} \textbf{\bibinfo{volume}{75}},
  \bibinfo{pages}{657} (\bibinfo{year}{2003}).

\bibitem[{\citenamefont{Maeno et~al.}(2012)\citenamefont{Maeno, Kittaka,
  Nomura, Yonezawa, and Ishida}}]{pap:Maeno2012}
\bibinfo{author}{\bibfnamefont{Y.}~\bibnamefont{Maeno}},
  \bibinfo{author}{\bibfnamefont{S.}~\bibnamefont{Kittaka}},
  \bibinfo{author}{\bibfnamefont{T.}~\bibnamefont{Nomura}},
  \bibinfo{author}{\bibfnamefont{S.}~\bibnamefont{Yonezawa}}, \bibnamefont{and}
  \bibinfo{author}{\bibfnamefont{K.}~\bibnamefont{Ishida}},
  \bibinfo{journal}{J.~Phys.~Soc.~Jpn.} \textbf{\bibinfo{volume}{81}},
  \bibinfo{pages}{011009} (\bibinfo{year}{2012}).

\bibitem[{\citenamefont{Ashby and Kallin}(2009)}]{pap:Ashby2009}
\bibinfo{author}{\bibfnamefont{P.~E.~C.} \bibnamefont{Ashby}} \bibnamefont{and}
  \bibinfo{author}{\bibfnamefont{C.}~\bibnamefont{Kallin}},
  \bibinfo{journal}{Phys.~Rev.~B} \textbf{\bibinfo{volume}{79}},
  \bibinfo{pages}{224509} (\bibinfo{year}{2009}).

\bibitem[{\citenamefont{Nagato et~al.}(2011)\citenamefont{Nagato, Higashitani,
  and Nagai}}]{pap:Nagato2011}
\bibinfo{author}{\bibfnamefont{Y.}~\bibnamefont{Nagato}},
  \bibinfo{author}{\bibfnamefont{S.}~\bibnamefont{Higashitani}},
  \bibnamefont{and} \bibinfo{author}{\bibfnamefont{K.}~\bibnamefont{Nagai}},
  \bibinfo{journal}{J.~Phys.~Soc.~Jpn.} \textbf{\bibinfo{volume}{80}},
  \bibinfo{pages}{113706} (\bibinfo{year}{2011}).

\bibitem[{\citenamefont{Read and Green}(2000)}]{pap:Read2000}
\bibinfo{author}{\bibfnamefont{N.}~\bibnamefont{Read}} \bibnamefont{and}
  \bibinfo{author}{\bibfnamefont{D.}~\bibnamefont{Green}},
  \bibinfo{journal}{Phys.~Rev.~B} \textbf{\bibinfo{volume}{61}},
  \bibinfo{pages}{10267} (\bibinfo{year}{2000}).

\bibitem[{\citenamefont{Luttinger}(1960)}]{pap:Luttinger1960}
\bibinfo{author}{\bibfnamefont{J.~M.} \bibnamefont{Luttinger}},
  \bibinfo{journal}{Phys.~Rev.} \textbf{\bibinfo{volume}{119}},
  \bibinfo{pages}{1153} (\bibinfo{year}{1960}).

\bibitem[{\citenamefont{Luttinger and Ward}(1960)}]{pap:LuttingerWard1960}
\bibinfo{author}{\bibfnamefont{J.~M.} \bibnamefont{Luttinger}}
  \bibnamefont{and} \bibinfo{author}{\bibfnamefont{J.~C.} \bibnamefont{Ward}},
  \bibinfo{journal}{Phys.~Rev.} \textbf{\bibinfo{volume}{118}},
  \bibinfo{pages}{1417} (\bibinfo{year}{1960}).

\bibitem[{\citenamefont{Yamanaka et~al.}(1997)\citenamefont{Yamanaka, Oshikawa,
  and Affleck}}]{pap:Oshikawa1997}
\bibinfo{author}{\bibfnamefont{M.}~\bibnamefont{Yamanaka}},
  \bibinfo{author}{\bibfnamefont{M.}~\bibnamefont{Oshikawa}}, \bibnamefont{and}
  \bibinfo{author}{\bibfnamefont{I.}~\bibnamefont{Affleck}},
  \bibinfo{journal}{Phys.~Rev.~Lett.} \textbf{\bibinfo{volume}{79}},
  \bibinfo{pages}{1110} (\bibinfo{year}{1997}).

\bibitem[{\citenamefont{Oshikawa}(2000)}]{pap:Oshikawa2000}
\bibinfo{author}{\bibfnamefont{M.}~\bibnamefont{Oshikawa}},
  \bibinfo{journal}{Phys.~Rev.~Lett.} \textbf{\bibinfo{volume}{84}},
  \bibinfo{pages}{3370} (\bibinfo{year}{2000}).

\bibitem[{\citenamefont{Dzyaloshinskii}(2003)}]{pap:Dzyaloshinskii2003}
\bibinfo{author}{\bibfnamefont{I.}~\bibnamefont{Dzyaloshinskii}},
  \bibinfo{journal}{Phys.~Rev.~B} \textbf{\bibinfo{volume}{68}},
  \bibinfo{pages}{085113} (\bibinfo{year}{2003}).

\bibitem[{com({\natexlab{a}})}]{com:j}
\bibinfo{note}{If contributions to surface currents from current density
  $\vecc{j}(x,y)$ were not localized around a surface, corresponding OM/OAM
  would not be proportional to area of a system since $\int dxdy
  (\vecc{r}\times \vecc{j})_z\sim O(L^{d+1})$ where $L$ is typical system
  length and $d=2$. Even if magnitude of $\vecc{j}$ is not exponentially
  decreasing and $\vecc{j}$ is oscillating away from a surface in a gapless
  system, contributions to surface currents from bulk regions would be
  negligible, because effects of a surface should be vanishingly small in the
  bulk regions and the bulk regions preserves the same translational symmetry
  as in the periodic boundary condition case. Besides, a surface region can be
  well identified when there is nearly uniform current density in the bulk
  which varies in a length scale much smaller than the system size. In this
  case, we can define the surface region so that the current density becomes
  nearly uniform (in the above sense) outside of it. Therefore, in general
  systems, there exists an appropriate surface region $S$ whose width is much
  smaller than the system size and surface currents are well defined by
  $I_y=\int_Sdx j_y$.}

\bibitem[{com({\natexlab{b}})}]{com:LW}
\bibinfo{note}{We note that the Luttinger-Ward indentity holds in the presence
  of both interactions and surface perturbations. For disorder surface
  perturbations, we sum up all the closed skeleton diagrams in terms of the
  averaged Green's function and construct a ``Luttinger-Ward functional''
  $\Phi[G]$ as in clean systems, whose derivative with respect to $G$ gives the
  averaged selfenergy. Then, we consider $\delta \Phi= \Phi[G+\delta
  G]-\Phi[G]$ with $\delta G(k)=\delta k_y \partial G(k)/\partial k_y$, which
  is a sum of all the closed skeleton diagrams but one $G$-line is replaced by
  $\delta G$-line for every diagram. Following the original argument for the
  Luttinger-Ward identity ~\cite{pap:Luttinger1960,pap:LuttingerWard1960}, it
  is now easily seen that $\sum_{k_y}{\rm tr}\Sigma \partial G/\partial k_y=0$
  holds due to the $k_y$-momentum conservation. By performing a partial
  integral, we obtain the Luttinger-Ward identity.}

\bibitem[{\citenamefont{Rosch}(2007)}]{pap:Rosch2007}
\bibinfo{author}{\bibfnamefont{A.}~\bibnamefont{Rosch}}, \bibinfo{journal}{Eur.
  Phys. J. B} \textbf{\bibinfo{volume}{59}}, \bibinfo{pages}{495}
  (\bibinfo{year}{2007}).

\bibitem[{\citenamefont{Dave et~al.}(2013)\citenamefont{Dave, Phillips, and
  Kane}}]{pap:Dave2013}
\bibinfo{author}{\bibfnamefont{K.~B.} \bibnamefont{Dave}},
  \bibinfo{author}{\bibfnamefont{P.~W.} \bibnamefont{Phillips}},
  \bibnamefont{and} \bibinfo{author}{\bibfnamefont{C.~L.} \bibnamefont{Kane}},
  \bibinfo{journal}{Phys.~Rev.~Lett.} \textbf{\bibinfo{volume}{110}},
  \bibinfo{pages}{090403} (\bibinfo{year}{2013}).

\bibitem[{\citenamefont{Landau}(1930)}]{pap:Landau1930}
\bibinfo{author}{\bibfnamefont{L.~D.} \bibnamefont{Landau}},
  \bibinfo{journal}{Z. Phys.} \textbf{\bibinfo{volume}{64}},
  \bibinfo{pages}{629} (\bibinfo{year}{1930}).

\bibitem[{\citenamefont{Peierls}(1933)}]{pap:Peierls1933}
\bibinfo{author}{\bibfnamefont{R.}~\bibnamefont{Peierls}}, \bibinfo{journal}{Z.
  Phys.} \textbf{\bibinfo{volume}{80}}, \bibinfo{pages}{763}
  (\bibinfo{year}{1933}).

\bibitem[{\citenamefont{Fukuyama}(1971)}]{pap:Fukuyama1971}
\bibinfo{author}{\bibfnamefont{H.}~\bibnamefont{Fukuyama}},
  \bibinfo{journal}{Prog.~Theor.~Phys.} \textbf{\bibinfo{volume}{45}},
  \bibinfo{pages}{704} (\bibinfo{year}{1971}).

\bibitem[{\citenamefont{Kubo}(1964)}]{pap:Kubo1964}
\bibinfo{author}{\bibfnamefont{R.}~\bibnamefont{Kubo}},
  \bibinfo{journal}{J.~Phys.~Soc.~Jpn.} \textbf{\bibinfo{volume}{19}},
  \bibinfo{pages}{2127} (\bibinfo{year}{1964}).

\bibitem[{\citenamefont{Ohtaka and Moriya}(1973)}]{pap:Ohtaka1973}
\bibinfo{author}{\bibfnamefont{K.}~\bibnamefont{Ohtaka}} \bibnamefont{and}
  \bibinfo{author}{\bibfnamefont{T.}~\bibnamefont{Moriya}},
  \bibinfo{journal}{J.~Phys.~Soc.~Jpn.} \textbf{\bibinfo{volume}{34}},
  \bibinfo{pages}{1203} (\bibinfo{year}{1973}).

\bibitem[{\citenamefont{Ishikawa and Fukuyama}(1999)}]{pap:Ishikawa1999}
\bibinfo{author}{\bibfnamefont{Y.}~\bibnamefont{Ishikawa}} \bibnamefont{and}
  \bibinfo{author}{\bibfnamefont{H.}~\bibnamefont{Fukuyama}},
  \bibinfo{journal}{J.~Phys.~Soc.~Jpn.} \textbf{\bibinfo{volume}{68}},
  \bibinfo{pages}{2405} (\bibinfo{year}{1999}).

\bibitem[{\citenamefont{Nee et~al.}(1968)\citenamefont{Nee, Koch, and
  Prange}}]{pap:Nee1968}
\bibinfo{author}{\bibfnamefont{T.~W.} \bibnamefont{Nee}},
  \bibinfo{author}{\bibfnamefont{J.~F.} \bibnamefont{Koch}}, \bibnamefont{and}
  \bibinfo{author}{\bibfnamefont{R.~E.} \bibnamefont{Prange}},
  \bibinfo{journal}{Phys.~Rev.} \textbf{\bibinfo{volume}{174}},
  \bibinfo{pages}{758} (\bibinfo{year}{1968}).

\bibitem[{\citenamefont{Abrikosov}(1988)}]{book:Abrikosov}
\bibinfo{author}{\bibfnamefont{A.~A.} \bibnamefont{Abrikosov}},
  \emph{\bibinfo{title}{Fundamentals of the Theory of Metals}}
  (\bibinfo{publisher}{North Holland}, \bibinfo{year}{1988}).

\bibitem[{\citenamefont{Bohm}(1949)}]{pap:Bohm1949}
\bibinfo{author}{\bibfnamefont{D.}~\bibnamefont{Bohm}},
  \bibinfo{journal}{Phys.~Rev.} \textbf{\bibinfo{volume}{75}},
  \bibinfo{pages}{502} (\bibinfo{year}{1949}).

\bibitem[{\citenamefont{Vignale}(1995)}]{pap:Vignale1995}
\bibinfo{author}{\bibfnamefont{G.}~\bibnamefont{Vignale}},
  \bibinfo{journal}{Phys.~Rev.~B} \textbf{\bibinfo{volume}{51}},
  \bibinfo{pages}{2612} (\bibinfo{year}{1995}).

\bibitem[{\citenamefont{Ohashi and Momoi}(1996)}]{pap:Ohashi1996}
\bibinfo{author}{\bibfnamefont{Y.}~\bibnamefont{Ohashi}} \bibnamefont{and}
  \bibinfo{author}{\bibfnamefont{T.}~\bibnamefont{Momoi}},
  \bibinfo{journal}{J.~Phys.~Soc.~Jpn.} \textbf{\bibinfo{volume}{65}},
  \bibinfo{pages}{3254} (\bibinfo{year}{1996}).

\bibitem[{\citenamefont{Kusama and Ohashi}(1999)}]{pap:Kusama1999}
\bibinfo{author}{\bibfnamefont{Y.}~\bibnamefont{Kusama}} \bibnamefont{and}
  \bibinfo{author}{\bibfnamefont{Y.}~\bibnamefont{Ohashi}},
  \bibinfo{journal}{J.~Phys.~Soc.~Jpn.} \textbf{\bibinfo{volume}{68}},
  \bibinfo{pages}{987} (\bibinfo{year}{1999}).

\bibitem[{\citenamefont{Lieb et~al.}(1961)\citenamefont{Lieb, Schultz, and
  Mattis}}]{pap:LSM1961}
\bibinfo{author}{\bibfnamefont{E.}~\bibnamefont{Lieb}},
  \bibinfo{author}{\bibfnamefont{T.}~\bibnamefont{Schultz}}, \bibnamefont{and}
  \bibinfo{author}{\bibfnamefont{D.}~\bibnamefont{Mattis}},
  \bibinfo{journal}{Ann. Phys.} \textbf{\bibinfo{volume}{16}},
  \bibinfo{pages}{407} (\bibinfo{year}{1961}).

\bibitem[{\citenamefont{Carruthers and Nieto}(1968)}]{pap:Carruthers1968}
\bibinfo{author}{\bibfnamefont{P.}~\bibnamefont{Carruthers}} \bibnamefont{and}
  \bibinfo{author}{\bibfnamefont{M.~M.} \bibnamefont{Nieto}},
  \bibinfo{journal}{Rev.~Mod.~Phys.} \textbf{\bibinfo{volume}{40}},
  \bibinfo{pages}{411} (\bibinfo{year}{1968}).

\bibitem[{\citenamefont{Emch}(1972)}]{book:Emch1972}
\bibinfo{author}{\bibfnamefont{G.~G.} \bibnamefont{Emch}}, in
  \emph{\bibinfo{booktitle}{Phase Transitions and Critical Phenomena}}
  (\bibinfo{publisher}{Academic Press, London}, \bibinfo{year}{1972}),
  vol.~\bibinfo{volume}{1}, chap.~\bibinfo{chapter}{4}.

\bibitem[{\citenamefont{Eschrig et~al.}(1985)\citenamefont{Eschrig, Seifert,
  and Ziesche}}]{pap:Eschrig1985}
\bibinfo{author}{\bibfnamefont{H.}~\bibnamefont{Eschrig}},
  \bibinfo{author}{\bibfnamefont{G.}~\bibnamefont{Seifert}}, \bibnamefont{and}
  \bibinfo{author}{\bibfnamefont{P.}~\bibnamefont{Ziesche}},
  \bibinfo{journal}{Solid State Commun.} \textbf{\bibinfo{volume}{56}},
  \bibinfo{pages}{777} (\bibinfo{year}{1985}).

\bibitem[{\citenamefont{Vignale and Rasolt}(1987)}]{pap:Vignale1987}
\bibinfo{author}{\bibfnamefont{G.}~\bibnamefont{Vignale}} \bibnamefont{and}
  \bibinfo{author}{\bibfnamefont{M.}~\bibnamefont{Rasolt}},
  \bibinfo{journal}{Phys.~Rev.~Lett.} \textbf{\bibinfo{volume}{59}},
  \bibinfo{pages}{2360} (\bibinfo{year}{1987}).

\bibitem[{\citenamefont{Higuchi and Hasegawa}(1997)}]{pap:Higuchi1997}
\bibinfo{author}{\bibfnamefont{M.}~\bibnamefont{Higuchi}} \bibnamefont{and}
  \bibinfo{author}{\bibfnamefont{A.}~\bibnamefont{Hasegawa}},
  \bibinfo{journal}{J.~Phys.~Soc.~Jpn.} \textbf{\bibinfo{volume}{66}},
  \bibinfo{pages}{149} (\bibinfo{year}{1997}).

\bibitem[{\citenamefont{Streda}(1982)}]{pap:Streda1982}
\bibinfo{author}{\bibfnamefont{P.}~\bibnamefont{Streda}}, \bibinfo{journal}{J.
  Phys. C} \textbf{\bibinfo{volume}{15}}, \bibinfo{pages}{L717}
  (\bibinfo{year}{1982}).

\bibitem[{\citenamefont{MacDonald}(1995)}]{book:MacDonald1995}
\bibinfo{author}{\bibfnamefont{A.~H.} \bibnamefont{MacDonald}}, in
  \emph{\bibinfo{booktitle}{Mesoscopic Quantum Physics (Les Houches, Session
  LXI)}} (\bibinfo{publisher}{North Holland, Amsterdam}, \bibinfo{year}{1995}).

\bibitem[{\citenamefont{Kaur et~al.}(2005)\citenamefont{Kaur, Agterberg, and
  Sigrist}}]{pap:Kaur2005}
\bibinfo{author}{\bibfnamefont{R.~P.} \bibnamefont{Kaur}},
  \bibinfo{author}{\bibfnamefont{D.~F.} \bibnamefont{Agterberg}},
  \bibnamefont{and} \bibinfo{author}{\bibfnamefont{M.}~\bibnamefont{Sigrist}},
  \bibinfo{journal}{Phys.~Rev.~Lett.} \textbf{\bibinfo{volume}{94}},
  \bibinfo{pages}{137002} (\bibinfo{year}{2005}).

\bibitem[{\citenamefont{Yip}(2005)}]{pap:Yip2005}
\bibinfo{author}{\bibfnamefont{S.~K.} \bibnamefont{Yip}}, \bibinfo{journal}{J.
  Low Temp. Phys.} \textbf{\bibinfo{volume}{140}}, \bibinfo{pages}{67}
  (\bibinfo{year}{2005}).

\bibitem[{\citenamefont{Eagles}(1969)}]{pap:Eagles1969}
\bibinfo{author}{\bibfnamefont{D.~M.} \bibnamefont{Eagles}},
  \bibinfo{journal}{Phys.~Rev.} \textbf{\bibinfo{volume}{186}},
  \bibinfo{pages}{456} (\bibinfo{year}{1969}).

\bibitem[{\citenamefont{Leggett}(1980)}]{book:Leggett1980}
\bibinfo{author}{\bibfnamefont{A.~J.} \bibnamefont{Leggett}}, in
  \emph{\bibinfo{booktitle}{Modern Trends in the Theory of Condensed Matter}},
  edited by \bibinfo{editor}{\bibfnamefont{A.}~\bibnamefont{Pekalski}}
  \bibnamefont{and} \bibinfo{editor}{\bibfnamefont{J.}~\bibnamefont{Przystawa}}
  (\bibinfo{publisher}{Springer, Berlin}, \bibinfo{year}{1980}).

\bibitem[{\citenamefont{Giorgini et~al.}(2008)\citenamefont{Giorgini,
  Pitaevskii, and Stringari}}]{pap:BCSBECreview2008}
\bibinfo{author}{\bibfnamefont{S.}~\bibnamefont{Giorgini}},
  \bibinfo{author}{\bibfnamefont{L.~P.} \bibnamefont{Pitaevskii}},
  \bibnamefont{and}
  \bibinfo{author}{\bibfnamefont{S.}~\bibnamefont{Stringari}},
  \bibinfo{journal}{Rev.~Mod.~Phys.} \textbf{\bibinfo{volume}{80}},
  \bibinfo{pages}{1215} (\bibinfo{year}{2008}).

\bibitem[{\citenamefont{Massignan et~al.}(2010)\citenamefont{Massignan,
  Sanpera, and Lewenstein}}]{pap:Massignan2010}
\bibinfo{author}{\bibfnamefont{P.}~\bibnamefont{Massignan}},
  \bibinfo{author}{\bibfnamefont{A.}~\bibnamefont{Sanpera}}, \bibnamefont{and}
  \bibinfo{author}{\bibfnamefont{M.}~\bibnamefont{Lewenstein}},
  \bibinfo{journal}{Phys.~Rev.~A} \textbf{\bibinfo{volume}{81}},
  \bibinfo{pages}{031607(R)} (\bibinfo{year}{2010}).

\bibitem[{com({\natexlab{c}})}]{com:Tsuruta}
\bibinfo{note}{A. Tsuruta and K. Miyake, unpublished.}

\end{thebibliography}

\end{document}